\newif\ifapj\apjfalse
\apjtrue

\ifapj
\documentclass[apj,numberedappendix]{emulateapj}
\usepackage{amsmath}
\else
\documentclass[12pt,preprint]{aastex}
\fi

\shorttitle{Reanalysis of CANGAROO-I data}
\shortauthors{Yoshikoshi et al.}

\begin{document}

\title{Reanalysis of Data Taken by the CANGAROO 3.8 Meter Imaging
Atmospheric Cherenkov Telescope: PSR~B1706--44, SN~1006, and Vela}

\author{
T.~Yoshikoshi\altaffilmark{1},
M.~Mori\altaffilmark{1,2},
P.~G.~Edwards\altaffilmark{3},
S.~Gunji\altaffilmark{4},
S.~Hara\altaffilmark{5},
T.~Hara\altaffilmark{6},
A.~Kawachi\altaffilmark{7},
Y.~Mizumoto\altaffilmark{8},
T.~Naito\altaffilmark{6},
K.~Nishijima\altaffilmark{7},
T.~Tanimori\altaffilmark{9},
G.~J.~Thornton\altaffilmark{10},
and T.~Yoshida\altaffilmark{11}
}

\email{tyoshiko@icrr.u-tokyo.ac.jp}

\altaffiltext{1}{Institute for Cosmic Ray Research, University of
Tokyo, Kashiwa, Chiba 277-8582, Japan.}
\altaffiltext{2}{Current address: Department of Physics, Ritsumeikan
University, Kusatsu, Shiga 525-8577, Japan.}
\altaffiltext{3}{Narrabri Observatory, Australia Telescope National
Facility, CSIRO, Narrabri, NSW 2390, Australia.}
\altaffiltext{4}{Department of Physics, Yamagata University, Yamagata,
Yamagata 990-8560, Japan.}
\altaffiltext{5}{Ibaraki Prefectural University of Health Sciences,
Ami, Ibaraki 300-0394, Japan.}
\altaffiltext{6}{Faculty of Management Information, Yamanashi Gakuin
University, Kofu, Yamanashi 400-8575, Japan.}
\altaffiltext{7}{Department of Physics, Tokai University, Hiratsuka,
Kanagawa 259-1292, Japan.}
\altaffiltext{8}{National Astronomical Observatory of Japan, Mitaka,
Tokyo 181-8588, Japan.}
\altaffiltext{9}{Department of Physics, Kyoto University, Sakyo-ku,
Kyoto 606-8502, Japan.}
\altaffiltext{10}{School of Chemistry and Physics, University of
Adelaide, SA 5005, Australia.}
\altaffiltext{11}{Faculty of Science, Ibaraki University, Mito,
Ibaraki 310-8512, Japan.}

\begin{abstract}

We have reanalyzed data from observations of PSR~B1706--44, SN~1006,
and the Vela pulsar region made with the CANGAROO 3.8\,m Imaging
Atmospheric Cherenkov Telescope between 1993 and 1998 in response to
the results reported for these sources by the H.E.S.S.\
Collaboration. Although detections of TeV gamma-ray emission from
these sources were claimed by CANGAROO more than 10 years ago, upper
limits to the TeV gamma-ray signals from PSR~B1706--44 and SN~1006
derived by H.E.S.S.\ are about an order of magnitude lower. The
H.E.S.S.\ group detected strong diffuse TeV gamma-ray emission from
Vela but with a morphology differing from the CANGAROO result.
In our reanalysis, in which gamma-ray selection criteria have been
determined exclusively using gamma-ray simulations and
\textsc{off}-source data as background samples, no significant TeV
gamma-ray signals have been detected from compact regions around
PSR~B1706--44 or within the northeast rim of SN~1006. The upper limits
to the integral gamma-ray fluxes at the 95\% confidence level have
been estimated for the 1993 data of PSR~B1706--44 to be $F(> 3.2 \pm
1.6\,\textrm{TeV}) < 8.03 \times
10^{-13}\,\textrm{photons}\;\textrm{cm}^{-2}\,\textrm{s}^{-1}$, for
the 1996 and 1997 data of SN~1006 to be $F(> 3.0 \pm
1.5\,\textrm{TeV}) < 1.20 \times
10^{-12}\,\textrm{photons}\;\textrm{cm}^{-2}\,\textrm{s}^{-1}$ and
$F(> 1.8 \pm 0.9\,\textrm{TeV}) < 1.96 \times
10^{-12}\,\textrm{photons}\;\textrm{cm}^{-2}\,\textrm{s}^{-1}$,
respectively. We discuss reasons why the original analyses gave the
source detections.
The reanalysis did result in a TeV gamma-ray signal from the Vela
pulsar region at the $4.5 \sigma$ level using 1993, 1994, and 1995
data. The excess was located at the same position, $0\fdg13$ to the
southeast of the Vela pulsar, as that reported in the original
analysis. We have investigated the effect of the acceptance
distribution in the field of view of the 3.8\,m telescope, which
rapidly decreases toward the edge of the field of the camera, on the
detected gamma-ray morphology. The expected excess distribution for
the 3.8\,m telescope has been obtained by reweighting the distribution
of HESS~J0835--455 measured by H.E.S.S.\ with the acceptance of the
3.8\,m telescope. The result is morphologically comparable to the
CANGAROO excess distribution, although the profile of the
acceptance-reweighted H.E.S.S.\ distribution is more diffuse than that
of CANGAROO. The integral gamma-ray flux from HESS~J0835--455 has been
estimated for the same region as defined by H.E.S.S.\ from the
1993--1995 data of CANGAROO to be $F(> 4.0 \pm 1.6\,\textrm{TeV}) =
(3.28 \pm 0.92) \times
10^{-12}\,\textrm{photons}\;\textrm{cm}^{-2}\,\textrm{s}^{-1}$, which
is statistically consistent with the integral flux obtained by
H.E.S.S.
\end{abstract}

\keywords{gamma rays: observations --- methods: data analysis ---
pulsars: individual (PSR~B1706--44, Vela pulsar) --- supernova
remnants --- supernovae: individual (SN~1006)}

\section{Introduction}

The CANGAROO\footnote{Acronym for the ``Collaboration of Australia and
Nippon (Japan) for a GAmma Ray Observatory in the Outback''.} 3.8\,m
telescope was an Imaging Atmospheric Cherenkov Telescope (IACT) for
TeV gamma-ray astrophysics which operated from 1992 to 1998 near
Woomera, South Australia. It is retrospectively called the CANGAROO-I
telescope since its pioneering role as the first IACT in the southern
hemisphere was subsequently inherited by the CANGAROO-II telescope
\citep{tanimori99} and the CANGAROO-III telescope system
\citep{kubo04}. The discoveries of TeV gamma-ray signals from five
sources have been claimed from CANGAROO-I data: PSR~B1706--44
\citep{kifune95}, the Vela pulsar region \citep{yoshikoshi97}, SN~1006
\citep{tanimori98}, RX~J1713.7--3946 \citep{muraishi00}, and
PSR~B1509--58\footnote{A marginal detection compared to the other
CANGAROO-I sources.} \citep{sako00}. However, H.E.S.S.\ has later
reported upper limits to the TeV gamma-ray fluxes from PSR~B1706--44
\citep{aharonian05a} and SN~1006 \citep{aharonian05b}, which are lower
than the CANGAROO-I fluxes by factors of $\sim 10$. The results are
obviously inconsistent as the reported statistical significances of
the CANGAROO-I detections were $12 \sigma$ for PSR~B1706--44 and $7.7
\sigma$ for SN~1006, assuming that the flux levels are constant over the
10 year period. More recently, H.E.S.S.\ has detected a TeV gamma-ray
signal from SN~1006 after accumulating more data
\citep{naumann-godo08}, but the flux level is well below their
previous upper
limit.
\footnote{\url{http://www.mpi-hd.mpg.de/hfm/HESS/pages/home/som/2008/08/}.}
The Durham group also reported a statistically significant ($5.9
\sigma$) detection of TeV gamma rays from PSR~B1706--44
\citep{chadwick98}. Preliminary support of the TeV gamma-ray signals
from PSR~B1706--44 and SN~1006 were reported from CANGAROO-II
observations \citep{kushida01, kushida03, hara01}.

H.E.S.S.\ also observed the Vela pulsar region in 2004 and 2005,
and detected a strong gamma-ray signal at the 50\% Crab level above
1\,TeV \citep{aharonian06a}. The emission is diffuse over the X-ray
(0.9--2.0\,keV) jet-like image, which was first detected by
\textit{ROSAT} \citep{markwardt95}, and extends about $1\degr$ from
the pulsar toward the south-southwest, reaching the position of the
brightest radio emission in the Vela supernova remnant (SNR), Vela~X
\citep{weiler80}. The position of the maximum TeV emission in
HESS~J0835--455 is located in the middle of the jet-like feature,
which has been identified by \textit{Chandra} to be ejected in the
direction of the equatorial pulsar wind \citep{helfand01,
pavlov03}. On the other hand, the TeV emission detected by CANGAROO-I
was offset from the pulsar by $0\fdg13$ to the southeast
\citep{yoshikoshi97}, corresponding to the possible pulsar birthplace
calculated using the proper motion of
$45\,\textrm{mas}\;\textrm{yr}^{-1}$ toward a position angle of
$301\degr$ \citep{dodson03} and the characteristic age of 11\,kyr
\citep{manchester05}. The CANGAROO-I source thus appears to be
different from the H.E.S.S.\ source, as the angular offset between
their peaks is $0\fdg34$, although the position of the CANGAROO-I peak 
does lie within the region of the H.E.S.S.\ diffuse emission.

The CANGAROO-I results for Vela obtained by \citet{yoshikoshi97} were
based on the data taken in 1993, 1994, and 1995. The Vela pulsar
region was also observed in 1997 with the recoated 3.8\,m reflector
and a TeV gamma-ray signal was again detected, at the $4.1 \sigma$
level \citep{yoshikoshi98}. The 1997 data were later also analyzed by
\citet{dazeley01b}, who reproduced the same signal as
\citet{yoshikoshi98} but cast doubt upon its gamma-ray-likeness. No
signal was found within a $2\degr$ field of view centered on the
pulsar position using gamma-ray selection criteria optimized \textit{a
priori} with their Monte Carlo simulations \citep{dazeley01a}.

In response to these inconsistencies, we have reanalyzed CANGAROO-I
data of PSR~B1706--44, SN~1006, and Vela, including the data used in
the previous analyses, with our best current knowledge. The following
facts were not known or not fully considered in the previous analyses:
(1) the effects of some noise produced by the electronic system were
only discovered after the previous publications; (2) details of the
analyses of CANGAROO-I data had varied from source to source, and
sometimes from data set to data set, without a common, consistent
treatment based on the experience gained over the lifetime of the
telescope; (3) improved calibration methods had not been applied to
some old data; (4) although the fact that the acceptance for gamma rays
is not uniform across the field of view was well understood, the
effect on gamma-ray morphologies had not been investigated in detail;
and (5) for Vela, a more precise energy spectrum, measured by H.E.S.S.,
is available now \citep{aharonian06a} and the systematic uncertainty
of the integral flux has been reduced. Our aim in this paper is to
investigate these issues on the CANGAROO-I data in detail, although
the information available on the previous analyses is, in places,
limited. We briefly summarize the CANGAROO 3.8\,m IACT and its
specifications related to the analyses in Section~\ref{sec_telescope},
and then the data of the three objects are summarized in
Section~\ref{sec_data}. The common and standard analysis method
redefined considering the above problems is described in
Section~\ref{sec_analyses} in detail, and the results obtained with it
are summarized in Section~\ref{sec_results}. Several issues on the old
results such as reproducibility and reliability are discussed in
Section~\ref{sec_discussion}, and we finally conclude in
Section~\ref{sec_conclusions}.

\section{CANGAROO 3.8\,m Telescope} \label{sec_telescope}

The CANGAROO 3.8\,m IACT \citep{hara93} was operated near Woomera,
South Australia ($136\degr47\arcmin$E, $31\degr06\arcmin$S, 160\,m
a.s.l.). The telescope detected Cherenkov photons from extensive air
showers (EASs) generated by primary gamma rays and cosmic rays. The
parabolic reflector of the telescope had both a diameter and a focal
length of 3.8\,m. The telescope was originally used for lunar ranging,
and the reflectivity of its Kanigen-plated surface was about 45\% at
wavelengths of atmospheric Cherenkov light. The reflector was recoated
with aluminum in 1996 October and its reflectivity improved to about
75\%,\footnote{This value is the average of the whole surface and
smaller than that of the recoated surface ($\sim 80$\%) since only the
outer part of the reflector was recoated.} which deteriorated to about
55\% by 1998. The reflectivity was monitored using a portable
reflectometer \citep{dowden97} with a systematic error of about 5\%.

The telescope had an imaging camera of Hamamatsu R2248 square
photomultiplier tubes (PMTs) in the focal plane. Observations started
with 224 PMTs, which were increased to 256 PMTs in 1995 April by adding
34 PMTs at the corners of the camera. The number of PMTs decreased to
240 because of hardware trouble in 1998 February. Each PMT viewed an
angular extent of $0\fdg12 \times 0\fdg12$, and the field of view
of the camera was about $3\degr$ across with a $0\fdg18$ spacing
between pixel centers. Light guides were used from 1995 November to
capture Cherenkov photons otherwise incident on the dead space between
photocathodes but abandoned in 1996 November owing to complications
described later.

Outputs of the PMTs were fed into the electronics, consisting of
trigger and data acquisition circuits after amplification by buffer
amplifiers. The data acquisition circuits are based on
analog-to-digital converters (ADCs), time-to-digital converters
(TDCs), and scalers for individual PMTs, which record brightness and
arrival timing of Cherenkov light, and brightness of night-sky
background (NSB) light, respectively.

\section{Data} \label{sec_data}

\begin{deluxetable}{clrrrr}
\tablecaption{Summary of the CANGAROO-I data sets used in this paper.
\label{tab_data}}
\tablewidth{0pt}
\tablehead{
\colhead{Target} & \colhead{Year} &
\multicolumn{2}{c}{$T_\textsc{on}$\tablenotemark{d} (hr)} &
\multicolumn{2}{c}{$T_\textsc{off}$\tablenotemark{d} (hr)}
}
\startdata
PSR B1706--44    & 1993\tablenotemark{a} & 66.9 & (44.3)    & 50.3 &
(44.3)    \\
                 & 1994                  & 19.4 & (19.1)    & 19.6 &
(19.1)    \\
                 & 1995                  & 33.9 & (26.0)    & 28.4 &
(26.0)    \\
                 & 1997                  & 24.1 & (18.4)    & 19.8 &
(18.4)    \\
                 & 1998                  & 44.2 & (31.5)    & 35.0 &
(31.5)    \\
Vela             & 1993\tablenotemark{b} & 52.0 & (32.6)    & 42.2 &
(32.6)    \\
                 & 1994\tablenotemark{b} & 63.8 & (41.9)    & 56.7 &
(41.9)    \\
                 & 1995\tablenotemark{b} & 50.2 & (44.9)    & 46.7 &
(44.9)    \\
                 & 1997                  & 28.8 & (16.6)    & 21.4 &
(16.6)    \\
SN 1006          & 1996\tablenotemark{c} & 26.8 & (\phn8.2) & 14.7 &
(\phn8.2) \\
                 & 1997\tablenotemark{c} & 34.7 & (28.4)    & 29.6 &
(28.4)    \\
\enddata
\tablenotetext{a}{Used in \citet{kifune95}.}
\tablenotetext{b}{Used in \citet{yoshikoshi97}.}
\tablenotetext{c}{Used in \citet{tanimori98}.}
\tablenotetext{d}{Observation time for \textsc{on}-source
(\textsc{off}-source) after rejecting periods affected by clouds or
significant electronic noise. The number in brackets is the
observation time after matching \textsc{on}- and \textsc{off}-source
observations in horizontal coordinates.}
\end{deluxetable}
The CANGAROO-I data used in this paper are listed in
Table~\ref{tab_data}. The data have been taken in
\textsc{on}/\textsc{off} mode, in which \textsc{off}-source data to
estimate the background level are taken on the same night, but with an
offset in right ascension, as the \textsc{on}-source observation which
included the target object in the field of view. The total observation
times for \textsc{on}- and \textsc{off}-source data in
Table~\ref{tab_data} ($T_\textsc{on}$ and $T_\textsc{off}$,
respectively) are those after rejecting periods affected by clouds or
significant electronic noise, which are determined from descriptions
in logbooks and/or background event rates. The \textsc{on}- and
\textsc{off}-source data have further been matched in horizontal
coordinates of the observations to make observation conditions as
similar as possible, since some data were not taken following the
above policy, and the corresponding observation times are given in
brackets. In the following analyses, we use the data sets after
matching \textsc{on}- and \textsc{off}-source exposures unless
otherwise specified.

The data used in the previous papers are indicated by the superscript
marks in the year column in Table~\ref{tab_data}. PSR~B1706--44 was
also observed in 1992, and those data were used in the analysis of
\cite{kifune95}. However, they are not used here because some of the
original data files have been lost and some necessary calibration data
were not taken at that time, and it is expected that the performance
over this first observing season will be more variable as problems
were identified and addressed.

\section{Analyses} \label{sec_analyses}

Except where otherwise specified, a common analysis method is used for
all data sets in this paper, since the nature of Cherenkov light images
of EASs is independent of the observed object. In the previous
CANGAROO-I papers, the analysis conditions differed for each object
without a single, consistent underlying philosophy, in part as the
understanding of the telescope improved with time. It is possible,
however, to enhance an apparent signal by using some of the many
degrees of freedom in reducing the background level. To reduce the
degrees of freedom in the analyses and ensure the results are robust,
we use here a simple analysis method similar to the ``Supercuts''
method of the Whipple group \citep{punch91}, which is described in
Section~\ref{sec_imaging}.

\subsection{Calibration of Tracking Centers} \label{sec_tracking}

\begin{figure}
\ifapj
\includegraphics[width=\linewidth]{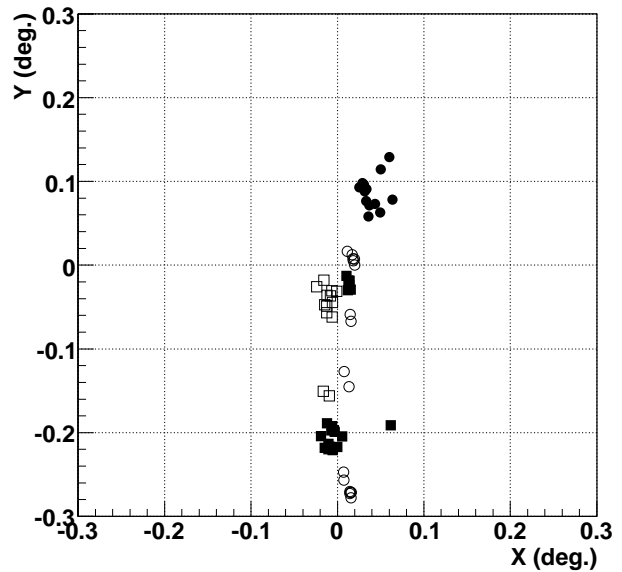}
\else
\plotone{f1.eps}
\fi
\caption{Calibrated tracking centers in the camera in the case of the Vela
observations over five years. The origin is the center of the camera,
and the Y-axis points to the zenith. Tracking centers of the 1993,
1994, 1995, and 1997 data are represented by solid circles, open
circles, solid squares, and open squares,
respectively. \label{fig_tracking_center}}
\end{figure}

The telescope tracking position in the field of the imaging camera is
calibrated using the time variation of scaler counts of individual
PMTs. When a bright star image is within the field of view of a PMT,
its count rate increases. Since the telescope has an alt-azimuth
mount, bright star images revolve around the center of the field of
view of the camera, and the time profile of scaler counts shows peaks
due to the passages of stars. The offset angle of the tracking
direction from the center of the camera can be estimated by comparing
the observed time profiles with simulated ones. It is necessary for
this calibration to be done night by night as issues with the mount
were found to sometimes result in offsets of greater than $0\fdg1$,
comparable to the size of the point-spread function (PSF).
Figure~\ref{fig_tracking_center} shows an example of tracking centers
calibrated night by night in the case of the Vela observations over five
years, in which the maximum difference between the vertical offsets is
more than $0\fdg4$.
Details of the calibration procedure are given in
\citet{yoshikoshi96}.

\subsection{Image Generation and Cleaning} \label{sec_image_cleaning}

The relative Cherenkov light signal of the $i$th pixel $s_i$ is
obtained using the following formula:
\begin{equation}
s_i = \frac{a_i - p_i}{g_i}, \label{eq_flatfield}
\end{equation}
where $a_i$, $p_i$, and $g_i$ are the ADC value, the ADC pedestal
value, and the relative gain of the $i$th pixel, respectively. The
ADC pedestal value of each pixel is determined using Cherenkov light
triggered events. After omitting any pixel which contains a signal or is
adjacent to a pixel containing a signal, the ADC distribution of each
pixel is fitted by a Gaussian distribution. The presence of a TDC
signal is used to determine that a pixel contains a signal. The ADC
pedestal value, taken to be the mean value of the Gaussian
distribution, is subtracted from the ADC value. The pedestal standard
deviation represents the noise level of the ADC value due to NSB light
and electronic noise. Any PMT signal lower than the $1 \sigma$ threshold
based on the noise level is rejected from the image. Pixel gains
relative to their average are determined using LED data, which were
taken by artificial triggers with a LED, permanently positioned at the
front of the camera for flashing uniform light at the all PMTs, before
or after Cherenkov light observation runs. Cherenkov light images are
flat-fielded, i.e., divided by $g_i$ as in equation~(\ref{eq_flatfield}).

Background levels of hadronic events estimated using
\textsc{off}-source data are biased by the difference of NSB
brightness between \textsc{on}- and \textsc{off}-sources. Software
padding \citep{cawley93} is an effective method to compensate for the
difference, as demonstrated successfully by the Whipple group. We have
tried to apply this method to some CANGAROO-I data sets, but did not
find it to be effective. This is due to the fact that ADC
signals have already been padded by electronic noise, which were
generated only when high voltages were supplied to the PMTs, just like
the original ``hardware'' padding utilized in most first-generation
ACT systems \citep{fruin68, weekes72, cawley93}. The other sources of
noise described in the following sections have larger effects and
parameter distributions of background events match reasonably well
between \textsc{on}- and \textsc{off}-source regions after considering
them. Therefore, software padding is not used in this paper.

In 1997, we found that the discriminators which precede the TDCs and
scalers generate noise when they are fired, significantly affecting
the ADCs on the same circuit board. This results in inherent offsets
in individual ADCs in such events. This ``ADC offset'' noise can
however be measured by using the ADC itself and by setting
discriminator thresholds to be very low compared to their normal
values (typically three photoelectrons). In the most affected channels,
the offsets amount to about 40 ADC counts, which correspond to about eight
photoelectrons and thus distorts dimmer Cherenkov images
significantly. We use the following formula to compensate for the ADC
offset $o_i$ instead of Equation~(\ref{eq_flatfield}):
\begin{equation}
s_i = \frac{a_i - p_i - o_i}{g_i}. \label{eq_adc_offset}
\end{equation}

From the pixels remaining after the above selections, isolated pixels,
which are not adjacent to any other pixels containing signals, are
further removed to extract clusters in the image. This process was
traditionally designed to remove effects of the NSB noise. The
remaining pixels are used to define the image parameters described in
Section~\ref{sec_imaging}.

\subsection{Rejection of Electronic Noise Events} \label{sec_noise}

The imaging camera consists of box units in which eight adjacent PMTs
selected to have similar gains share the same high voltage. In early
data, in particular, noise events in which all eight PMTs in a box
unit were simultaneously hit were frequently generated by the
telescope drive system. These noise events, referred to as ``box
noise'', must be rejected as much as possible because they are compact
and mimic gamma-ray events. We define a parameter $b$ to reject box
noise as follows:
\begin{equation}
b = \frac{\max\limits_i n_i}{n}, \label{eq_box_noise}
\end{equation}
where $n_i$ is the number of TDC hits in the $i$th box of eight PMTs and
$n$ is the total number of TDC hits in an image. The box noise events
have large values of $b$ close to (or equal to) its maximum value of
1. In the standard analysis here, events of $b$ greater than 0.8 are
rejected. This rejection is very effective unless two or more boxes
are triggered in this way simultaneously, although small Cherenkov
images are also somewhat reduced as a sacrifice. About 15\% of
reconstructed events are rejected with this cut on average, but the
fraction can be larger depending on the noise level.

Using signals of the selected pixels, the image \textit{size} (sum of
the ADC values), the number of selected pixels (referred to as
$N_\mathrm{hit}$ though it is not the true number of TDC hits), and
the image centroid (first moment of the image) are calculated. Events
of small \textit{size} or $N_\mathrm{hit}$ are more affected by noise,
and therefore events of $\textit{size} < 200$ ADC counts or
$N_\mathrm{hit} < 5$ are rejected before the following imaging
analysis. Also, images of the centroids more than $1\fdg05$ distant
from the center of the camera are eliminated. This is called the
``edge cut'' since part of such an image has possibly been lost out of
the camera edge, resulting in the image shape being distorted. On average, the
\textit{size}, $N_\mathrm{hit}$, and edge cuts individually reject
about 40\%, 25\%, and 55\% of reconstructed events, respectively, and
about 75\% of reconstructed events are rejected by the three cuts
applied all together.

\subsection{Imaging Analysis} \label{sec_imaging}

The atmospheric Cherenkov imaging analysis is based on the Hillas
parameters \citep{hillas85, weekes89, reynolds93}, which are the
second moments of light intensity both parallel (\textit{length}) and
perpendicular (\textit{width}) to the major axis of the image, the
light concentration of the two brightest pixels (\textit{conc}), the
distance of the image centroid from the source (\textit{dis}), and the
orientation angle with respect to the major axis
(\textit{alpha}). They are also referred to as shape (\textit{width},
\textit{length}, and \textit{conc}), location (\textit{dis}), and
orientation (\textit{alpha}) parameters. Images of gamma-ray-induced
air showers tend to have smaller \textit{width},
\textit{length}, and \textit{alpha}, and larger \textit{conc} than
background hadronic showers, reflecting the different profiles of
shower development and different source distributions in the field of
view. In the case of a point source,  the signal-to-noise ratio is
improved by more than an order of magnitude by applying selections for
gamma-ray-like events to the parameter values.

The standard gamma-ray selection criteria used in this paper are
simple and represented as follows:
\begin{equation}
\begin{array}{ccccc}
               &      & \textit{width}  & \leq & w_\mathrm{max} \\
               &      & \textit{length} & \leq & l_\mathrm{max} \\
c_\mathrm{min} & \leq & \textit{conc}   &      &                \\
d_\mathrm{min} & \leq & \textit{dis}    & \leq & d_\mathrm{max} \\
               &      & \textit{alpha}  & \leq & a_\mathrm{max}.
\end{array}
\label{eq_selections}
\end{equation}
The best combination of the upper and lower limits in the above
selection criteria is determined so as to maximize the quality factor
$q$:
\begin{equation}
q = \frac{\epsilon_\gamma}{\sqrt{\epsilon_\mathrm{BG}}}, \label{eq_q}
\end{equation}
where $\epsilon_\gamma$ and $\epsilon_\mathrm{BG}$ are the
efficiencies of the selection criteria for gamma-ray and background
events, respectively. Gamma-ray event samples are generated using
Monte Carlo simulations, and for background event samples actual
\textsc{off}-source data are used. Only shape parameters,
\textit{width}, \textit{length}, and \textit{conc}, are used in the
optimization because the distributions of location and orientation
parameters depend on the source distribution that is not known
\textit{a priori}. Nevertheless, the best selection criteria on the
shape parameters depend on the source position in the camera,
especially on the offset angle of the source from the camera center,
which can change night by night owing to the shift of the tracking
center. The average offset angle is calculated in each data set and
incorporated in the Monte Carlo simulations.

\begin{figure}
\ifapj
\includegraphics[width=\linewidth]{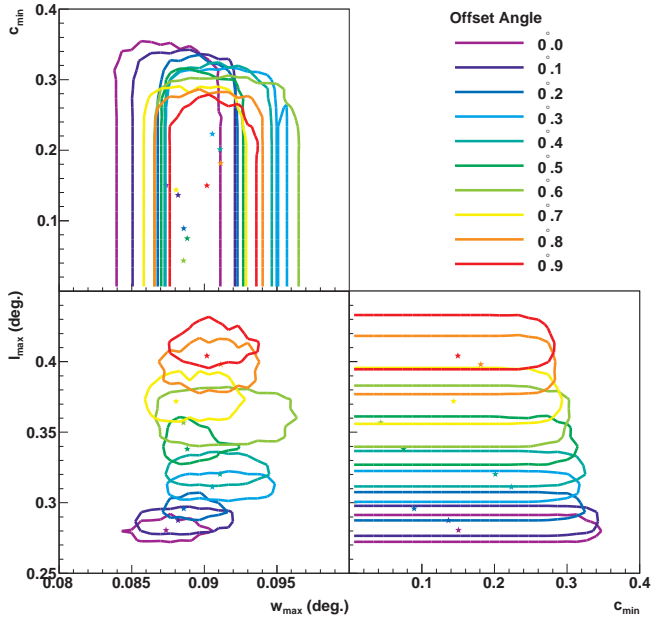}
\else
\plotone{f2.eps}
\fi
\caption{Correlations between the best gamma-ray selection criteria
giving the maximum $q$ factors in the case of the Vela 1993--1995
analysis. Bottom left: $l_\mathrm{max}$ (upper limit for
\textit{length}) vs. $w_\mathrm{max}$ (upper limit for \textit{width}),
top left: $c_\mathrm{min}$ (lower limit for \textit{conc}) vs.
$w_\mathrm{max}$, and bottom right: $c_\mathrm{min}$ vs.
$l_\mathrm{max}$. The star marks indicate the best combinations of the
selection criteria and the lines indicate their $1 \sigma$ error
regions. The different colors indicate different offset angles of the
source from the center of the camera with the $0\fdg1$
interval. \label{fig_best_q}}
\end{figure}

Figure~\ref{fig_best_q} shows correlations between the best gamma-ray
selection levels on the shape parameters with the \textsc{off}-source
data of the Vela 1993, 1994, and 1995 observations used as the
background samples.
The best combination of $w_\mathrm{max}$, $l_\mathrm{max}$, and
$c_\mathrm{min}$ for each offset angle of a point source from the
camera center is indicated by a star mark, and the closed line around
it represents the $1 \sigma$ error region, which is defined so that $q
\geq q_\mathrm{max} - \sigma_{q_\mathrm{max}}$. It is clearly seen
that the more distant the source position is from the camera center,
the larger the upper limit to \textit{length} $l_\mathrm{max}$ is, but
$w_\mathrm{max}$ and $c_\mathrm{min}$ are less dependent on the source
offset. The relative errors of $c_\mathrm{min}$ are larger than those
of $w_\mathrm{max}$ and $l_\mathrm{max}$ since \textit{conc} is
correlated with \textit{width} and \textit{length}. The positions
of the best shape selection levels indicated by the star marks seem
quite widely spread within the $1 \sigma$ error regions, especially for
\textit{conc}, since $q$ is a discrete function of the selection
levels by definition.

\ifapj
\begin{deluxetable*}{cccrrrr}
\tabletypesize{\scriptsize}
\else
\begin{deluxetable}{cccrrrr}
\tabletypesize{\small}
\fi
\tablecaption{Best gamma-ray selection criteria for each
data set. \label{tab_best_q}}
\tablewidth{0pt}
\tablehead{
\colhead{Target} & \colhead{Year} & \colhead{Offset\tablenotemark{a}}
& \colhead{$q_\mathrm{max}$} & \colhead{$w_\mathrm{max}$} &
\colhead{$l_\mathrm{max}$} & \colhead{$c_\mathrm{min}$}
}
\startdata
PSR B1706--44    & 1993             & $0\fdg1$ & $1.549 \pm 0.012$
& $0\fdg0883^{+0\fdg0046}_{-0\fdg0036}$ &
$0\fdg288^{+0\fdg013}_{-0\fdg007}$ & $0.17^{+0.18}_{-0.17}$ \\
                 & 1994             & $0\fdg1$ & $1.567 \pm 0.015$
& $0\fdg0873^{+0\fdg0057}_{-0\fdg0033}$ &
$0\fdg288^{+0\fdg025}_{-0\fdg008}$ & $0.34^{+0.03}_{-0.34}$ \\
                 & 1995             & $0\fdg1$ & $1.563 \pm 0.014$
& $0\fdg0896^{+0\fdg0026}_{-0\fdg0049}$ &
$0\fdg292^{+0\fdg024}_{-0\fdg012}$ & $0.15^{+0.20}_{-0.15}$ \\
                 & 1997             & $0\fdg1$ & $1.748 \pm 0.016$
& $0\fdg0856^{+0\fdg0052}_{-0\fdg0058}$ &
$0\fdg294^{+0\fdg017}_{-0\fdg011}$ & $0.08^{+0.27}_{-0.08}$ \\
                 & 1998             & $0\fdg1$ & $1.995 \pm 0.025$
& $0\fdg0810^{+0\fdg0047}_{-0\fdg0044}$ &
$0\fdg300^{+0\fdg015}_{-0\fdg026}$ & $0.12^{+0.25}_{-0.12}$ \\
Vela             & 1993, 1994, 1995 & $0\fdg2$ & $1.560 \pm 0.007$
& $0\fdg0886^{+0\fdg0034}_{-0\fdg0017}$ &
$0\fdg296^{+0\fdg012}_{-0\fdg009}$ & $0.09^{+0.25}_{-0.09}$ \\
                 &                  & $0\fdg4$ & $1.466 \pm 0.006$
& $0\fdg0911^{+0\fdg0035}_{-0\fdg0040}$ &
$0\fdg320^{+0\fdg016}_{-0\fdg010}$ & $0.20^{+0.13}_{-0.20}$ \\
                 & 1997             & $0\fdg1$ & $1.635 \pm 0.008$
& $0\fdg0929^{+0\fdg0037}_{-0\fdg0066}$ &
$0\fdg296^{+0\fdg016}_{-0\fdg009}$ & $0.23^{+0.10}_{-0.23}$ \\
                 &                  & $0\fdg4$ & $1.482 \pm 0.007$
& $0\fdg0953^{+0\fdg0036}_{-0\fdg0058}$ &
$0\fdg331^{+0\fdg028}_{-0\fdg014}$ & $0.10^{+0.22}_{-0.10}$ \\
SN 1006          & 1996             & $0\fdg3$ & $1.557 \pm 0.022$
& $0\fdg0962^{+0\fdg0032}_{-0\fdg0033}$ &
$0\fdg309^{+0\fdg025}_{-0\fdg012}$ & $0.23^{+0.08}_{-0.23}$ \\
                 &                  & $0\fdg7$ & $1.413 \pm 0.018$
& $0\fdg0957^{+0\fdg0029}_{-0\fdg0041}$ &
$0\fdg393^{+0\fdg028}_{-0\fdg065}$ & $0.15^{+0.13}_{-0.15}$ \\
                 & 1997             & $0\fdg3$ & $1.555 \pm 0.010$
& $0\fdg0924^{+0\fdg0030}_{-0\fdg0063}$ &
$0\fdg321^{+0\fdg014}_{-0\fdg021}$ & $0.13^{+0.21}_{-0.13}$ \\
\enddata
\tablenotetext{a}{Offset angle of the source from the center of the
camera.}
\ifapj
\end{deluxetable*}
\else
\end{deluxetable}
\fi
The best shape selection levels for individual data sets are listed in
Table~\ref{tab_best_q}. For each data set, $5 \times 10^5$ gamma-ray
showers have been simulated, randomly injected into a circular area of
300\,m radius around the telescope. Gamma-ray energies are also
randomly selected in the range between 500\,GeV and 100\,TeV assuming
that they follow a power-law differential spectrum or that with an
exponential cutoff. The assumed spectral index and cutoff energy are
listed in Table~\ref{tab_simulation} together with the other input
conditions in the simulations described in
Section~\ref{sec_telescope}. The spectral index and cutoff energy for
Vela are now known and taken from the reference given in the table,
but the spectral indices for PSR~B1706--44 and SN~1006 are assumed to
be the Crab-like value of $-2.5$. The zenith angles of gamma-ray
injections are selected near the culminations of the objects at which
most events were observed. In spite of the various differences in the
observation conditions, the best selection levels are consistent
within the errors, of which the definition is the same as in
Figure~\ref{fig_best_q}, except for the effect of the source offset
from the camera center.
\ifapj
\begin{deluxetable*}{cclccccc}
\else
\begin{deluxetable}{ccrrrrrr}
\fi
\tabletypesize{\scriptsize}
\tablecaption{Input conditions in the Monte Carlo simulations for each
data set. \label{tab_simulation}}
\tablewidth{0pt}
\tablecolumns{2}
\tablehead{
\colhead{Target} & \colhead{Year} & \colhead{Spectral Index} &
\colhead{\shortstack{Cutoff Energy\\(TeV)}} &
\colhead{\shortstack{Zenith Angle\\(deg)}} & \colhead{Light Guide} &
\colhead{\shortstack{Reflectivity\\(\%)}} & \colhead{Number of PMTs}
}
\startdata
PSR B1706--44    & 1993             & $-2.5$                   &
\nodata               & 15 & no  & 45 & 224 \\
                 & 1994             & $-2.5$                   &
\nodata               & 15 & no  & 45 & 224 \\
                 & 1995             & $-2.5$                   &
\nodata               & 15 & no  & 45 & 256 \\
                 & 1997             & $-2.5$                   &
\nodata               & 15 & no  & 75 & 256 \\
                 & 1998             & $-2.5$                   &
\nodata               & 15 & no  & 55 & 240 \\
Vela             & 1993, 1994, 1995 & $-1.45$\tablenotemark{a} &
13.8\tablenotemark{a} & 15 & no  & 45 & 224 \\
                 & 1997             & $-1.45$\tablenotemark{a} &
13.8\tablenotemark{a} & 15 & no  & 75 & 256 \\
SN 1006          & 1996             & $-2.5$                   &
\nodata               & 10 & yes & 45 & 256 \\
                 & 1997             & $-2.5$                   &
\nodata               & 10 & no  & 75 & 256 \\
\enddata
\tablenotetext{a}{\citet{aharonian06a}.}
\ifapj
\end{deluxetable*}
\else
\end{deluxetable}
\clearpage
\fi

The optimization of selection levels on location and orientation
parameters is complicated if the effect of source extent is
considered. Instead, we here take just simple criteria: the
\textit{dis} limits are common as $d_\mathrm{min} = 0\fdg7$ and
$d_\mathrm{max} = 1\fdg2$ since their dependence on the source
extent is not so strong, and $a_\mathrm{max} = 10\degr$ for sources
whose extent is comparable to the PSF, and for sources more extended
the same ranges as used in the previous CANGAROO-I paper
(Tanimori et al.\ 1998; $\textit{alpha} \leq 15\degr$)
are adopted.

\subsection{Estimation of Integral Gamma-Ray Fluxes} \label{sec_flux}

The integral gamma-ray fluxes obtained in this paper have been estimated
using the following formula:
\begin{equation}
F(> E_\mathrm{th}) = \frac{N_\mathrm{excess}}{N_\mathrm{expected}}
F_0(> E_\mathrm{th}),
\end{equation}
where $E_\mathrm{th}$ is the threshold energy for gamma rays, which is
defined as the modal energy of selected gamma rays, $N_\mathrm{excess}$
is the number of excess events, $F_0$ is the assumed integral
gamma-ray flux from the source, and $N_\mathrm{expected}$ is the
expected number of gamma rays represented as follows:
\begin{equation}
N_\mathrm{expected} = A_0 \epsilon t
\int_{E_\mathrm{min}}^{E_\mathrm{max}} f_0(E) dE,
\end{equation}
where $f_0 = - dF_0 / dE$ is the assumed differential gamma-ray flux
from the source, $E_\mathrm{min}$ and $E_\mathrm{max}$ are the minimum
and maximum energies of gamma rays generated in the simulations,
respectively, $A_0 = \pi r_\mathrm{max}^2 \cos \theta$ is the area
within which simulated gamma rays were injected, with the maximum
horizontal distance between the telescope and the shower axis
$r_\mathrm{max} = 300$\,m and the zenith angle $\theta$, $\epsilon$ is
the acceptance defined as $\epsilon = N_\mathrm{surviving} /
N_\mathrm{input}$, in which $N_\mathrm{input}$ and
$N_\mathrm{surviving}$ are the numbers of input gamma rays and gamma
rays surviving after the selections, respectively, and $t$ is the
observation time.

\section{Results} \label{sec_results}

\ifapj
\begin{deluxetable*}{cccrrrrrrr}
\else
\begin{deluxetable}{cccrrrrrrr}
\fi
\tabletypesize{\scriptsize}
\tablecaption{Summary of numbers of events remaining after various
selections. \label{tab_events}}
\tablewidth{0pt}
\tablehead{
\colhead{Target} & \colhead{Year} & \colhead{\textsc{on}/\textsc{off}}
& \colhead{Raw\tablenotemark{a}} & \colhead{Noise\tablenotemark{b}} &
\colhead{Shape\tablenotemark{c}} & \colhead{Location\tablenotemark{d}}
& \colhead{Orientation\tablenotemark{e} ($a_\mathrm{max}$)} &
\colhead{Excess} & \colhead{Significance}
}
\startdata
PSR B1706--44    & 1993             & \textsc{on}                   &
 180105 &  57587 & 14942 &  8669 &  1090 ($10\degr$)
                   & $-67$ &  $-1.4 \sigma$ \\
                 &                  & \textsc{off}                  &
 167397 &  56085 & 14570 &  8469 &  1157 ($10\degr$)
                   &       &                \\
                 & 1994             & \textsc{on}                   &
  62616 &  20602 &  5014 &  2763 &   366 ($10\degr$)
                   & $-24$ &  $-0.9 \sigma$ \\
                 &                  & \textsc{off}                  &
  63818 &  21186 &  5094 &  2863 &   390 ($10\degr$)
                   &       &                \\
                 & 1995             & \textsc{on}                   &
 159487 &  19865 &  4454 &  2445 &   299 ($10\degr$)
                   & $-31$ &  $-1.2 \sigma$ \\
                 &                  & \textsc{off}                  &
 138384 &  20508 &  4775 &  2664 &   330 ($10\degr$)
                   &       &                \\
                 & 1997             & \textsc{on}                   &
 114862 &  22541 &  3736 &  1994 &   230 ($10\degr$)
                   &  $-2$ &  $-0.1 \sigma$ \\
                 &                  & \textsc{off}                  &
 112291 &  22651 &  3792 &  2062 &   232 ($10\degr$)
                   &       &                \\
                 & 1998             & \textsc{on}                   &
  91547 &  18662 &  2422 &  1280 &   160 ($10\degr$)
                   &    22 &   $1.3 \sigma$ \\
                 &                  & \textsc{off}                  &
  84319 &  16972 &  2082 &  1090 &   138 ($10\degr$)
                   &       &                \\
Vela             & 1993, 1994, 1995 & \textsc{on}\tablenotemark{f}  &
 420310 & 119974 & 29710 & 16970 &  2343 ($10\degr$)
                   &   298 &   $4.5 \sigma$ \\
                 &                  & \textsc{off}\tablenotemark{f} &
 400418 & 120206 & 28867 & 16512 &  2045 ($10\degr$)
                   &       &                \\
                 &                  & \textsc{on}\tablenotemark{g}  &
 420310 & 119974 & 35190 &       & 10122\tablenotemark{j}
 \phm{($10\degr$)} &   501 &   $3.6 \sigma$ \\
                 &                  & \textsc{off}\tablenotemark{g} &
 400418 & 120206 & 34243 &       &  9621\tablenotemark{j}
 \phm{($10\degr$)} &       &                \\
                 & 1997             & \textsc{on}\tablenotemark{f}  &
 142158 &  36064 &  8039 &  4421 &   589 ($10\degr$)
                   &    84 &   $2.5 \sigma$ \\
                 &                  & \textsc{off}\tablenotemark{f} &
 132171 &  35292 &  8004 &  4309 &   505 ($10\degr$)
                   &       &                \\
                 &                  & \textsc{on}\tablenotemark{g}  &
 142158 &  36064 &  9862 &       &  2624\tablenotemark{j}
 \phm{($10\degr$)} &    75 &   $1.0 \sigma$ \\
                 &                  & \textsc{off}\tablenotemark{g} &
 132171 &  35292 &  9772 &       &  2549\tablenotemark{j}
 \phm{($10\degr$)} &       &                \\
SN 1006          & 1996             & \textsc{on}\tablenotemark{h}  &
  33524 &   2105 &   564 &   216 &    32 ($15\degr$)
                   &  $-5$ &  $-0.6 \sigma$ \\
                 &                  & \textsc{off}\tablenotemark{h} &
  27171 &   2122 &   581 &   213 &    37 ($15\degr$)
                   &       &                \\
                 &                  & \textsc{on}\tablenotemark{i}  &
  33524 &   2105 &   696 &   274 &    46 ($15\degr$)
                   &     2 &   $0.2 \sigma$ \\
                 &                  & \textsc{off}\tablenotemark{i} &
  27171 &   2122 &   728 &   266 &    44 ($15\degr$)
                   &       &                \\
                 & 1997             & \textsc{on}\tablenotemark{h}  &
 231398 &  57143 & 13572 &  6915 &  1246 ($15\degr$)
                   & $-94$ &  $-1.8 \sigma$ \\
                 &                  & \textsc{off}\tablenotemark{h} &
 226916 &  57112 & 14016 &  7199 &  1340 ($15\degr$)
                   &       &                \\
\enddata
\tablenotetext{a}{Number of reconstructed events (at least two
adjacent pixels containing signals).}
\tablenotetext{b}{Number of events remaining after the noise
rejection described in Section~\ref{sec_noise} and the discharge
noise rejection described in Section~\ref{sec_1006} for the SN~1006
1996 data.}
\tablenotetext{c}{Number of events remaining after the shape
selections based on \textit{width}, \textit{length}, and \textit{conc}.}
\tablenotetext{d}{Number of events remaining after the location
selection based on \textit{dis}.}
\tablenotetext{e}{Number of events remaining after the orientation
selection based on \textit{alpha}.}
\tablenotetext{f}{Analyzed with respect to the CANGAROO-I position
($\alpha = 8^\mathrm{h} 35^\mathrm{m} 42^\mathrm{s}$, $\delta =
-45\degr 17\arcmin$ (J2000)).}
\tablenotetext{g}{Analyzed with respect to HESS~J0835--455 ($\alpha =
8^\mathrm{h} 35^\mathrm{m} 00^\mathrm{s}$, $\delta = -45\degr
36\arcmin$ (J2000)).}
\tablenotetext{h}{Analyzed with respect to the NE rim with the
selection criteria optimized for the source offset by $0\fdg3$ from
the camera center.}
\tablenotetext{i}{Analyzed with respect to the NE rim with the
selection criteria optimized for the source offset by $0\fdg7$ from
the camera center.}
\tablenotetext{j}{Selected so that the arrival direction obtained
using the source PDF method \citep{yoshikoshi96} is within the angular
distance of $0\fdg8$ from HESS~J0835--455.}
\ifapj
\end{deluxetable*}
\else
\end{deluxetable}
\fi
The numbers of events remaining after the selections described above
are summarized in Table~\ref{tab_events} for all data sets used in this
paper. The selections listed in the table header have cumulatively
been applied to each data set from left to right. The statistical
significances of the excess events obtained after applying all
selections have been calculated using the formula obtained by
\citet{li83} utilizing the likelihood ratio. Note that after the noise
rejections the numbers of \textsc{on}- and \textsc{off}-source events
match with each other within a few percent except for the 1998
PSR~B1706--44 data. Details are given in the following subsections.

\subsection{PSR~B1706--44}

The five data sets of PSR~B1706--44 have been reanalyzed separately
since the 1993 data used in the previous CANGAROO-I paper should
independently be analyzed for comparison, and the mirror
reflectivities and camera configurations of the other data sets listed
in Table~\ref{tab_data} all differ. The average offset of the object
from the camera center is about $0\fdg1$, which has been assumed to
be common for all data sets. The gamma-ray selection criteria listed in
Table~\ref{tab_best_q} are optimized for this common offset, but the
selection levels are different from data set to data set since the
\textsc{off}-source data of each data set have been used as the
background samples in each optimization.

\begin{figure}
\ifapj
\includegraphics[width=\linewidth]{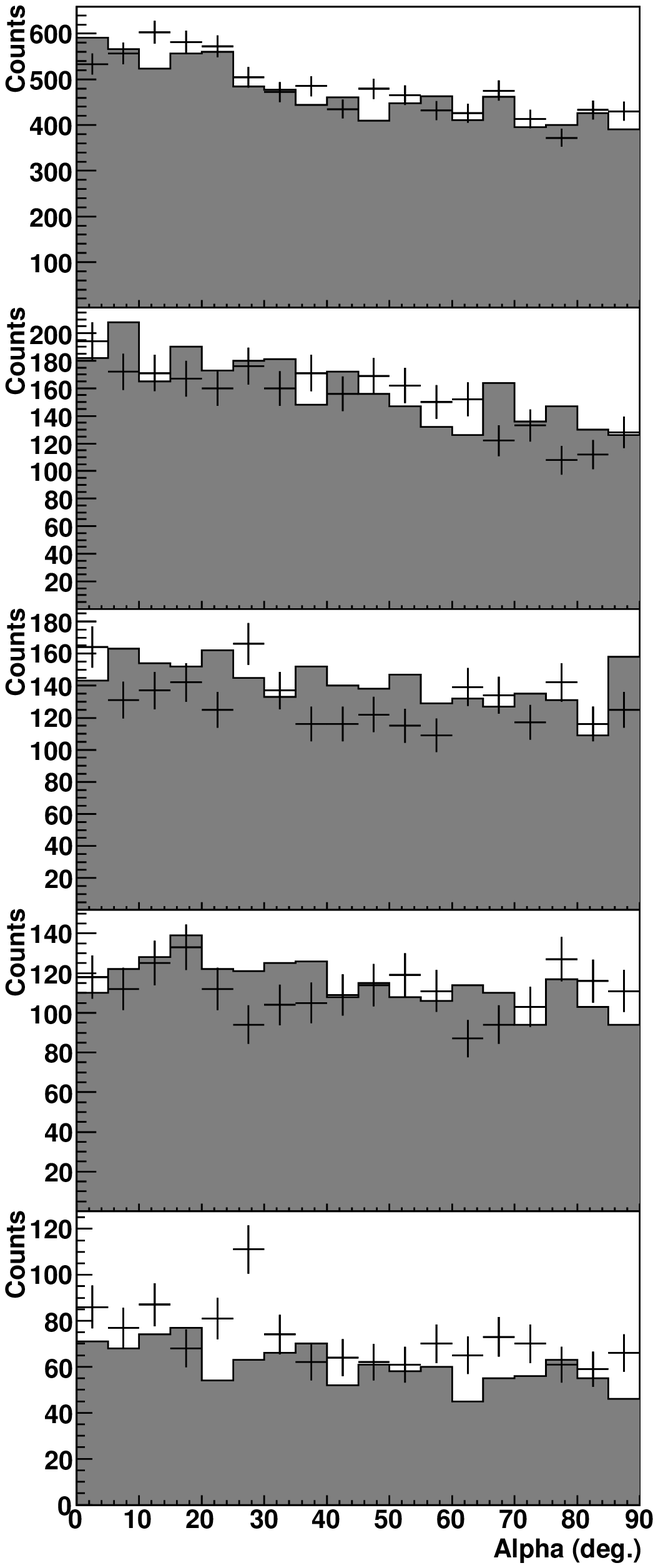}
\else
\epsscale{0.5}
\plotone{f3.eps}
\epsscale{1}
\fi
\caption{Distributions of the orientation parameter \textit{alpha} for
the PSR~B1706--44 data sets. From the top: distributions of the 1993,
1994, 1995, 1997, and 1998 data. The histograms with error bars are
the \textsc{on}-source distributions, and the gray solid histograms are
the \textsc{off}-source distributions. \label{fig_alpha_1706}}
\end{figure}

\begin{figure}
\ifapj
\includegraphics[width=\linewidth]{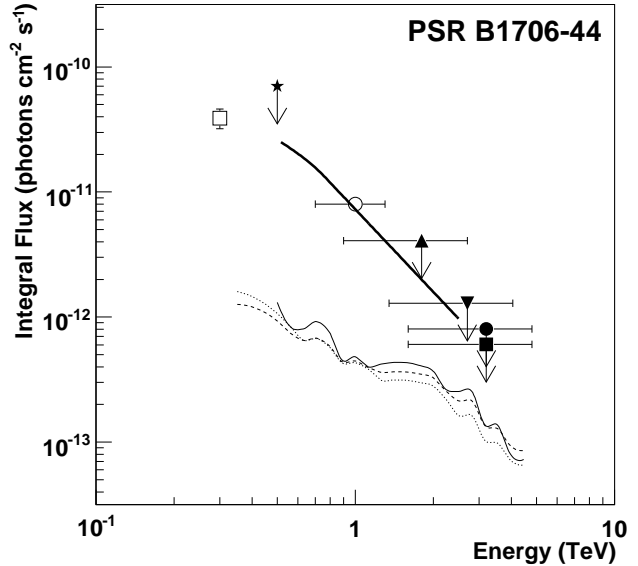}
\else
\plotone{f4.eps}
\fi
\caption{Integral gamma-ray fluxes from PSR~B1706--44 obtained using
the CANGAROO-I data. The 95\% CL upper limits to the integral flux
from PSR~B1706--44 obtained from the 1993, $1993 + 1994$, 1997, and
1998 data are indicated by the filled circle, filled square, filled
triangle, and filled upside-down triangle, respectively. The open
circle indicates the integral flux estimated by \citet{kifune95} using
the 1993 data. The horizontal bars represent the systematic errors of
the threshold energies. The 99\% CL upper limit curves obtained by
H.E.S.S.\ \citep{aharonian05a} are indicated by the thin solid,
dashed, and dotted lines. The thick solid line is the integral
spectrum converted from the preliminary CANGAROO-II result
\citep{kushida03}. The open square and filled star indicate the
results from \citet{chadwick98} and \citet{rowell98}, respectively.
\label{fig_flux_1706}}
\end{figure}

The \textit{alpha} distributions of \textsc{on}- and
\textsc{off}-source events remaining after the standard selections
have been plotted for individual data sets in
Figure~\ref{fig_alpha_1706}.
No statistically significant excess of \textsc{on}-source events has
been found near $\textit{alpha} = 0\degr$ in any data set, and the
significant excess from the 1993 data previously reported has not been
reproduced in this analysis. The \textsc{on}- and \textsc{off}-source
\textit{alpha} distributions of each data set agree well except for the
1998 data, in which the \textsc{on}-source distribution is somewhat
higher than the \textsc{off}-source distribution in the whole range. A
possible reason for this is a temporary decrease of the mirror
reflectivity due to dew condensation since it was humid in this period
and more \textsc{off}-source data was taken in the hours before
dawn. This diffuse excess is not gamma-ray-like considering the
distributions of the shape parameters of the excess events, and a
conservative upper limit is calculated here leaving the background
mismatch as it is. Setting $a_\mathrm{max} = 10\degr$ for a relatively
point-like source and assuming a Crab-like spectral index of $-2.5$ in
the simulations, the upper limits to the integral fluxes at the 95\%
confidence level (CL) have been estimated using the method described
in Section~\ref{sec_flux} and the method of
\citet{protheroe84}. The results are as follows:
\begin{displaymath}
F_{93}(> 3.2 \pm 1.6\,\textrm{TeV}) < 8.03 \times
10^{-13}\,\textrm{photons}\;\textrm{cm}^{-2}\,\textrm{s}^{-1},
\end{displaymath}
\begin{displaymath}
F_\textrm{\scriptsize 93--94}(> 3.2 \pm 1.6\,\textrm{TeV}) < 6.06
\times 10^{-13}\,\textrm{photons}\;\textrm{cm}^{-2}\,\textrm{s}^{-1},
\end{displaymath}
\begin{displaymath}
F_{95}(> 3.2 \pm 1.6\,\textrm{TeV}) < 8.85 \times
10^{-13}\,\textrm{photons}\;\textrm{cm}^{-2}\,\textrm{s}^{-1},
\end{displaymath}
\begin{displaymath}
F_{97}(> 1.8 \pm 0.9\,\textrm{TeV}) < 4.08 \times
10^{-12}\,\textrm{photons}\;\textrm{cm}^{-2}\,\textrm{s}^{-1},
\end{displaymath}
\begin{displaymath}
F_{98}(> 2.7 \pm 1.4\,\textrm{TeV}) < 1.29 \times
10^{-12}\,\textrm{photons}\;\textrm{cm}^{-2}\,\textrm{s}^{-1},
\end{displaymath}
where $F_\textrm{\scriptsize 93--94}$ is the integral flux calculated
adding the 1993 and 1994 data sets in which the observation conditions
are the same. The different threshold energies for the above upper
limits are mostly due to the different mirror reflectivities listed in
Table~\ref{tab_simulation}. The errors of the threshold energies are
systematic errors, defined in the same way as described by
\citet{muraishi00}. In the calculation of the flux upper limit
of the 1997 data, only 7.9\,hr of the \textsc{on}- and
\textsc{off}-source data, which corresponds to 43\% of the data used
elsewhere, have been used since blue optical filters were tested in
the rest of the observations and their effect has not accurately been
determined. The upper limits are plotted in
Figure~\ref{fig_flux_1706}.
They are higher than, and therefore not inconsistent with, the
H.E.S.S.\ upper limits reported by \citet{aharonian05a}.

\subsection{SN~1006} \label{sec_1006}

The SN~1006 data taken in 1996 and 1997 have been analyzed with
respect to the northeast (NE) rim ($\alpha = 15^\mathrm{h}
03^\mathrm{m} 54^\mathrm{s}$, $\delta = -41\degr 45\arcmin 30\arcsec$
(J2000)), from which \citet{tanimori98} have claimed the detection of
TeV gamma rays. The observations were made with average offsets of
$0\fdg3$ (June 1996 and 1997) and $0\fdg7$ (1996 April) of the NE
rim from the camera center in order to move the bright star $\beta$
Lup (magnitude 2.68) out of the field of view of the camera.
Because of these relatively large offsets, the best gamma-ray
selection criteria for SN~1006 listed in Table~\ref{tab_best_q} are
looser than the others following the tendency indicated in
Figure~\ref{fig_best_q}. In the case of the 1996 data, the looser
criteria are also caused by the light guides\footnote{Light guides are
utilized to detect Cherenkov photons otherwise incident on the gaps
between photocathodes of PMTs but at the same time serve to somewhat
smear Cherenkov images.} which were used only in this period. In the
case of CANGAROO-I, the collection efficiency of the light guides is
thought to be worse than originally expected owing to difficulties in
placing them in proximity to the photocathodes which resulted in
discharge noise generated between the light guides and photocathodes,
and the Cherenkov event rate in fact decreased after their
installation. Here, to generate simulation events, we assume the light
guide efficiency to be 45\%, which is almost the same as the filling
factor of the photocathodes. As in the case of PSR~B1706--44, the
Crab-like spectral index $-2.5$ is assumed in the simulations.

As mentioned above, the 1996 data were contaminated by discharge
noise, in which the hardware trigger rate suddenly increased,
especially in humid conditions, with only one or two PMTs having very
large ADC values in one image. Such images are compact and have large
sizes, and therefore, mimic gamma-ray images. We here take the
following simple method to reject the discharge noise events.
\begin{enumerate}
\item For each pixel, the rate in which the pixel has the maximum ADC
value in the image is calculated.
\item Pixels with rates higher than 0.01\,Hz are marked as noisy
pixels night by night (the rate of normal pixels is about 0.001\,Hz).
\item Discharge noise events are identified and rejected if the pixel
having the maximum ADC value in the image coincides with one of the
marked pixels for the night.
\end{enumerate}
This discharge noise cut is more robust than the old method
described in Section~\ref{sec_old_1006} and more than 75\% of shower
events remain after this cut. The numbers of \textsc{on}- and
\textsc{off}-source events after the cut match quite well and the
difference between them is only 0.8\%.

The 1997 data are free from the discharge noise since the light guides
were removed in 1996 November. However, the number of
\textsc{on}-source events after the other noise cuts described in
Section~\ref{sec_noise} is more than a few percent larger than that of
\textsc{off}-source events. This mismatch is probably due to the
difference of brightness in the field of view between \textsc{on}- and
\textsc{off}-sources since another bright star, $\kappa$ Cen
(magnitude 3.13), was inevitably included in the \textsc{on}-source
field of view. To reduce the effect of bright stars, pixels having
scaler values greater than $100\,\textrm{counts}\;\textrm{ms}^{-1}$
were removed from the image in the analysis of the 1997 data. As a
result, the numbers of \textsc{on}- and \textsc{off}-source events
match quite well as shown in Table~\ref{tab_events}.

\begin{figure}
\ifapj
\includegraphics[width=\linewidth]{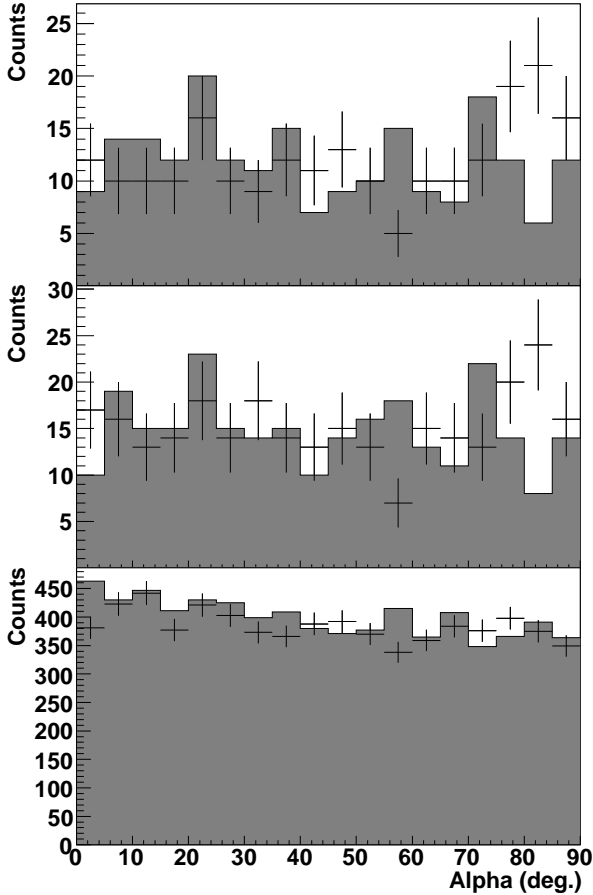}
\else
\epsscale{0.5}
\plotone{f5.eps}
\epsscale{1}
\fi
\caption{Distributions of the orientation parameter \textit{alpha} for
the SN~1006 data sets. The \textit{alpha} values have been calculated
with respect to the NE rim of the SNR. Top: distributions of the 1996
data with the best selection criteria optimized for the $0\fdg3$
offset from the camera center, middle: distributions of the 1996 data
with the best selection criteria for the $0\fdg7$ offset, and
bottom: distributions of the 1997 data with the best selection
criteria for the $0\fdg3$ offset. The histograms with the error bars
and the gray solid histograms are \textsc{on}- and \textsc{off}-source
distributions, respectively. \label{fig_alpha_1006}}
\end{figure}

\begin{figure}
\ifapj
\includegraphics[width=\linewidth]{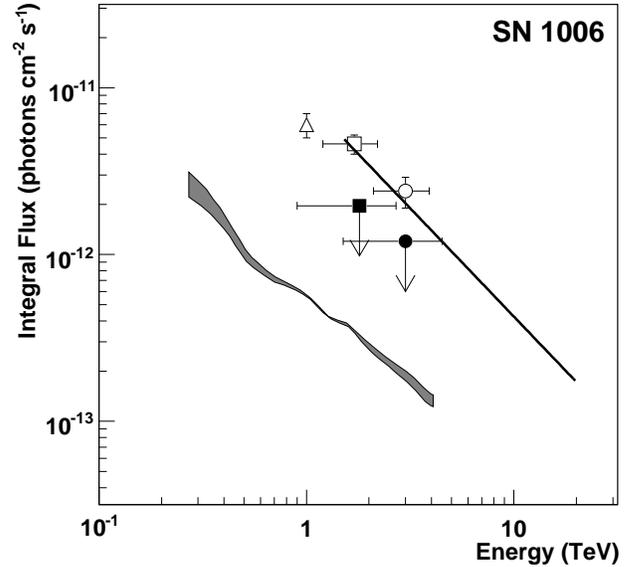}
\else
\plotone{f6.eps}
\fi
\caption{Integral gamma-ray fluxes from the NE rim of SN~1006 obtained
using the CANGAROO-I data. The 95\% CL upper limits to the integral
fluxes are indicated by the filled circle (1996 data with the
selection criteria optimized for the $0\fdg3$ offset) and the filled
square (1997 data). The open circle and open square are the integral
fluxes estimated by \citet{tanimori98} using the 1996 and 1997 data,
respectively. The thick solid line is the integral spectrum converted
from the CANGAROO-I differential spectrum \citep{tanimori01}. The
horizontal bars represent the systematic errors of the threshold
energies. The 99.9\% CL upper limits obtained by H.E.S.S.\
\citep{aharonian05b} are indicated by the gray band for a range of
assumed photon indices (2--3). The open triangle indicates the
preliminary CANGAROO-II result \citep{hara01}. \label{fig_flux_1006}}
\end{figure}

Figure~\ref{fig_alpha_1006} shows the \textit{alpha} distributions of
the SN~1006 data sets.
No statistically significant gamma-ray signal has been found around
$\textit{alpha} = 0\degr$ from the 1996 and 1997 data for both
$0\fdg3$ and $0\fdg7$ offsets from the camera center. Therefore,
the previous CANGAROO-I results have not been reproduced in this
analysis. Setting $a_\mathrm{max} = 15\degr$, which is the same
definition as used by \citet{tanimori98}, the 95\% CL upper limits to
the integral flux have been calculated as follows:
\begin{displaymath}
F_{96,0\fdg3}(> 3.0 \pm 1.5\,\textrm{TeV}) < 1.20 \times
10^{-12}\,\textrm{photons}\;\textrm{cm}^{-2}\,\textrm{s}^{-1},
\end{displaymath}
\begin{displaymath}
F_{96,0\fdg7}(> 3.0 \pm 1.5\,\textrm{TeV}) < 2.44 \times
10^{-12}\,\textrm{photons}\;\textrm{cm}^{-2}\,\textrm{s}^{-1},
\end{displaymath}
\begin{displaymath}
F_{97,0\fdg3}(> 1.8 \pm 0.9\,\textrm{TeV}) < 1.96 \times
10^{-12}\,\textrm{photons}\;\textrm{cm}^{-2}\,\textrm{s}^{-1}.
\end{displaymath}
The difference between the threshold energies is mostly due to the
different mirror reflectivities listed in
Table~\ref{tab_simulation}. The errors of the threshold energies are
systematic errors, as defined by
\citet{muraishi00}. These upper limits are plotted in
Figure~\ref{fig_flux_1006}.
They are higher than and not inconsistent with the H.E.S.S.\ upper
limits on the SN~1006 flux \citep{aharonian05b}.

\subsection{Vela} \label{sec_vela}

The analysis procedure has been applied to the Vela data, first with
respect to the CANGAROO-I position ($\alpha = 8^\mathrm{h}
35^\mathrm{m} 42^\mathrm{s}$, $\delta = -45\degr 17\arcmin$ (J2000)),
which is $0\fdg13$ offset from the Vela pulsar to the southeast
\citep{yoshikoshi97}. The 1993--1995 data and 1997 data have been
analyzed separately since the mirror reflectivity was improved in
1996. The average offsets of the object from the camera center are
about $0\fdg2$ (1993--1995) and $0\fdg1$ (1997), for which the
shape selection criteria listed in Table~\ref{tab_best_q} have been
optimized. The obtained \textit{alpha} distributions are shown in
Figure~\ref{fig_alpha_vela}.
\ifapj
\begin{figure*}
\includegraphics[width=0.5\linewidth]{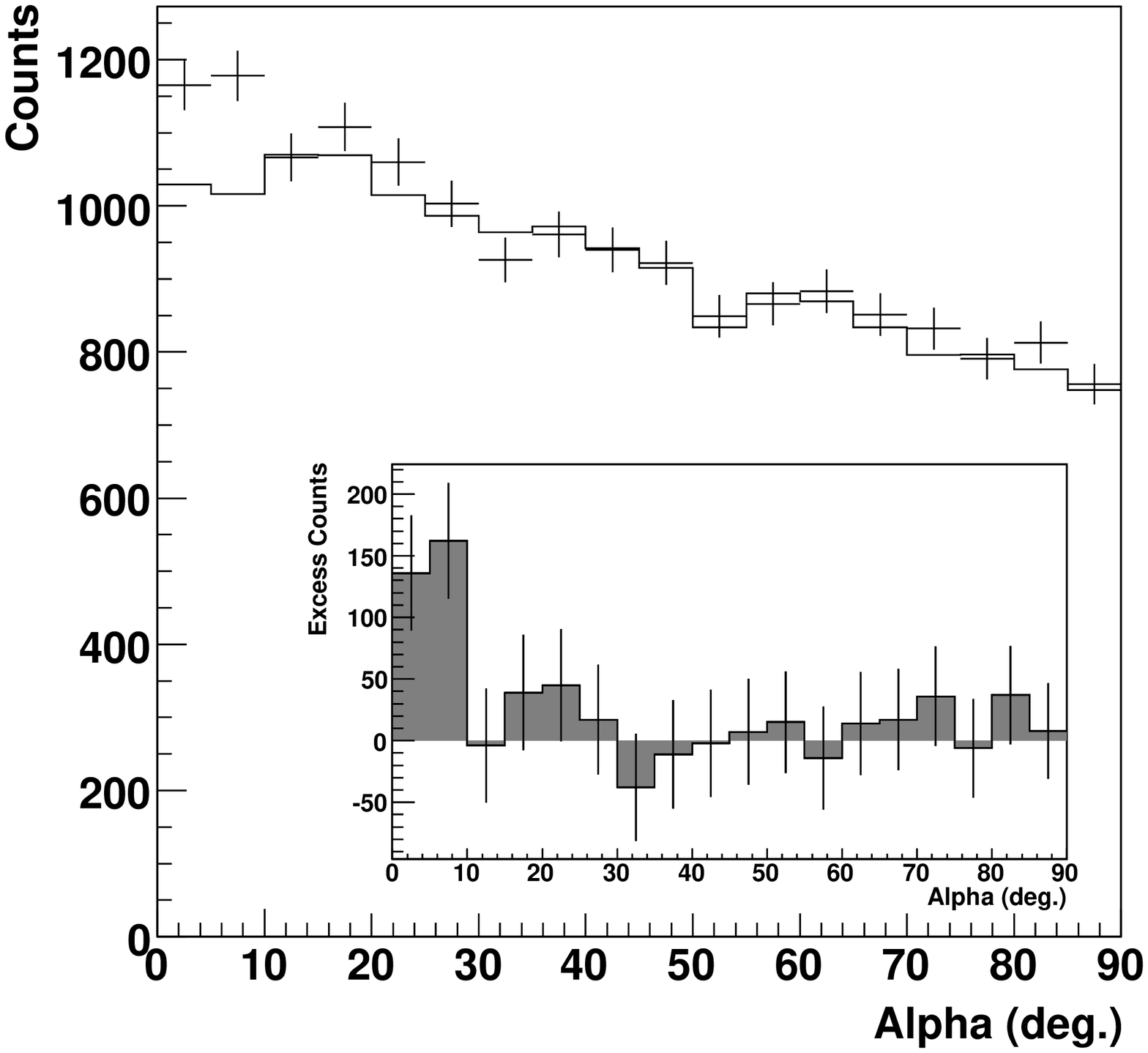}
\includegraphics[width=0.5\linewidth]{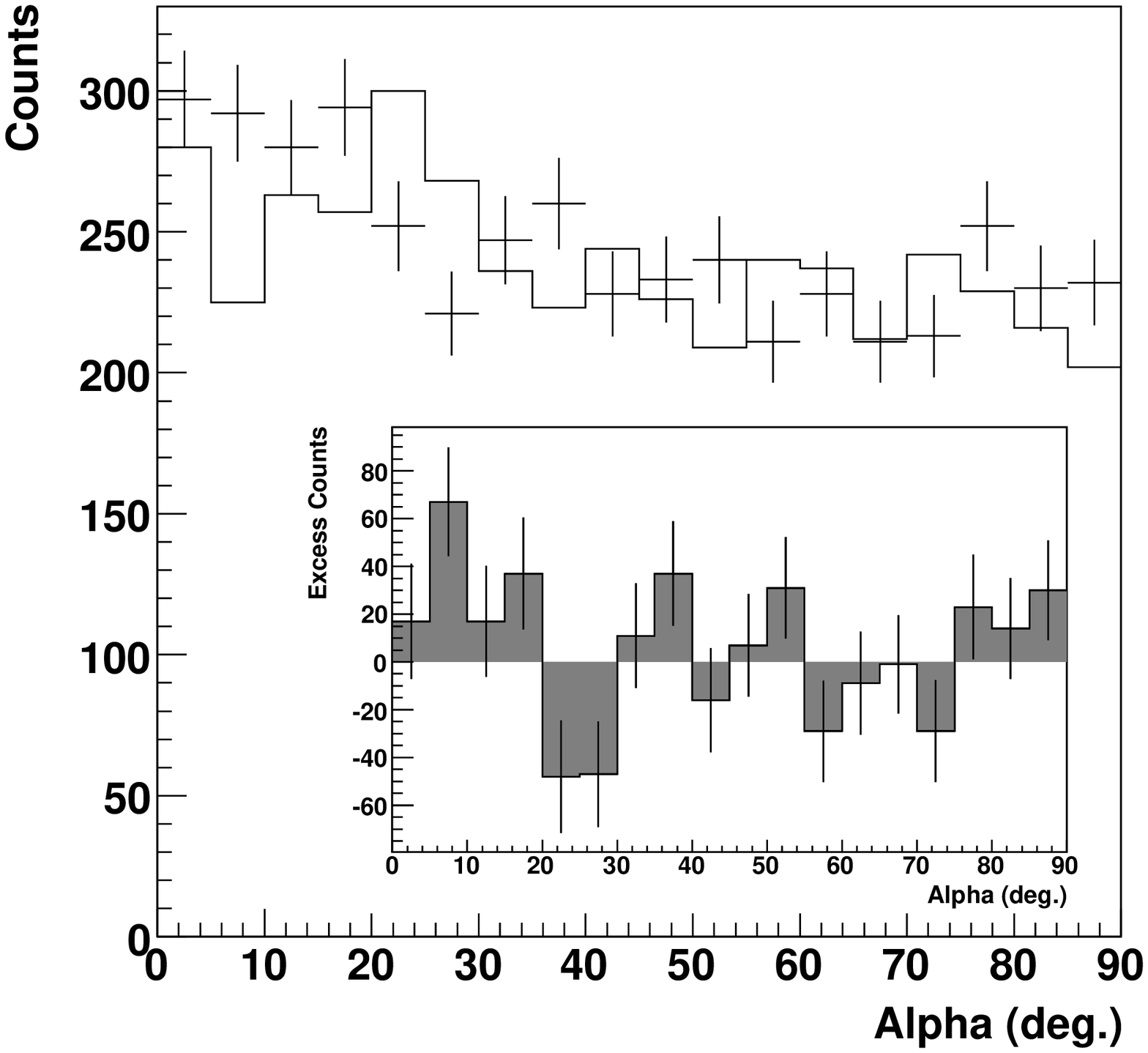}
\else
\begin{figure}
\plottwo{f7a.eps}{f7b.eps}
\fi
\caption{Distributions of the orientation parameter
\textit{alpha} for the Vela data. The \textit{alpha} values have been
calculated with respect to the CANGAROO-I position that is offset by
$0\fdg13$ from the Vela pulsar to the southeast. The left and right
figures have been obtained from the 1993--1995 data and the 1997
data, respectively. The \textsc{on}- and \textsc{off}-source
distributions are represented by the histograms with the error bars
and the solid histograms, respectively. The histograms in the insets
are the distributions of excess counts. \label{fig_alpha_vela}}
\ifapj
\end{figure*}
\else
\end{figure}
\fi
Setting $a_\mathrm{max} = 10\degr$, the statistical significances of
the 1993--1995 data and 1997 data are $4.5 \sigma$ and $2.5 \sigma$,
respectively. The former is smaller than the previous result of $5.2
\sigma$ \citep{yoshikoshi97}, but the signal is still at a significant
level. Note, however, that the CANGAROO-I position was found in the
previous analysis after searching for the highest peak in the $2\degr
\times 2\degr$ area around the Vela pulsar \citep{yoshikoshi97}, and
therefore the above values are ``pre-trial'' statistical
significances.

Considering the $0\fdg18$ PSF (half width at half maximum) of
CANGAROO-I \citep{yoshikoshi97}, the CANGAROO-I position does not
coincide with the peak position of HESS~J0835--455 detected by
H.E.S.S., which is shown at the top left of
Figure~\ref{fig_morphology_vela} \citep{aharonian06a}.
\ifapj
\begin{figure*}
\includegraphics[width=\textwidth]{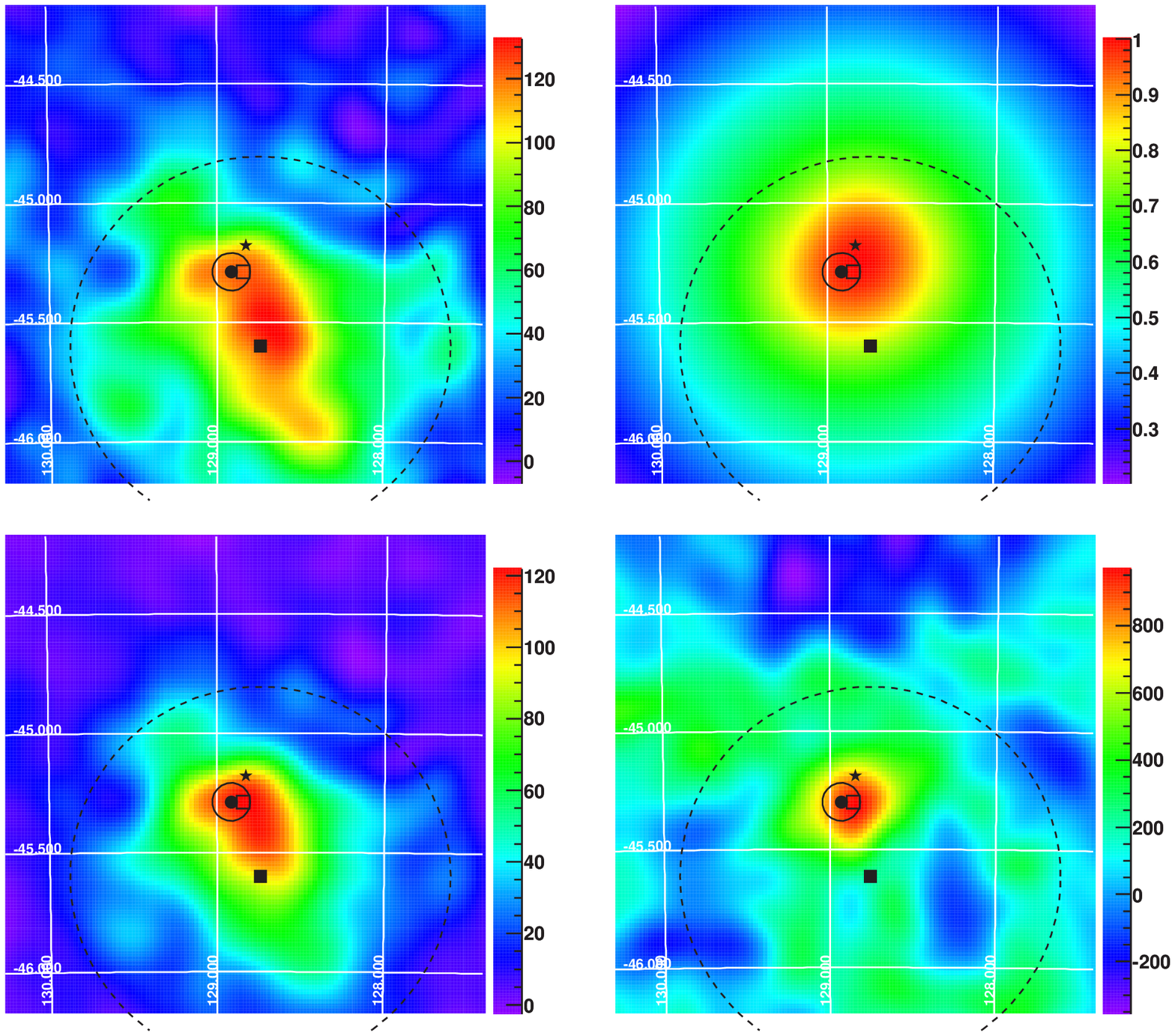}
\else
\begin{figure}
\plotone{f8.eps}
\fi
\caption{Sky maps of the Vela pulsar region. Top left: distribution of
excess counts obtained by H.E.S.S.\ with the threshold energy of
450\,GeV \citep{aharonian06a}, top right: averaged acceptance
distribution in the CANGAROO-I observations from 1993 to 1995
normalized so that the maximum value is equal to 1, bottom left:
H.E.S.S.\ excess distribution reweighted according to the CANGAROO-I
acceptance, where the H.E.S.S.\ acceptance \citep{aharonian06b} has
not been considered but is almost uniform within the field of view
compared to the CANGAROO-I acceptance, bottom right: distribution of
excess counts obtained by CANGAROO-I with the threshold energy of
4\,TeV using the source PDF method \citep{yoshikoshi96}, in which the
PSF size is $0\fdg18$ (half width at half maximum). The rainbow
scales at the right are in counts per $0\fdg02 \times 0\fdg02$
area for the H.E.S.S.\ distributions and in
$\textrm{counts}\;\textrm{deg}^{-2}$ for the CANGAROO
distribution. The star mark at the center of each sky map indicates
the position of the Vela pulsar. The filled circle indicates the peak
position of the CANGAROO-I excess, whereas the filled and open squares
indicate the position of the center of gravity of the H.E.S.S.\ excess
and the peak position of the acceptance-reweighted H.E.S.S.\ excess,
respectively. The small open circle around the CANGAROO-I position
represents the $2 \sigma$ location error ($1 \sigma \sim
0\fdg04$). The big dashed circle of the $0\fdg8$ radius indicates
the integration region for HESS~J0835--455 defined by H.E.S.S.\
\citep{aharonian06a}. \label{fig_morphology_vela}}
\ifapj
\end{figure*}
\else
\end{figure}
\fi
However, the acceptance $\epsilon$ is not uniform over the field of
view and the morphology could significantly be distorted by this
non-uniformity, especially for the CANGAROO-I imaging camera which had
a relatively small field of view, about $3\degr$ across. The top right
of Figure~\ref{fig_morphology_vela} is the averaged acceptance
distribution in the Vela pulsar region for the 1993--1995
observation period, obtained considering the night-by-night shifts of
the tracking center in the camera and rotation of the camera with
respect to the sky view due to the alt-azimuth mount. The expected
morphology for CANGAROO-I is obtained from the H.E.S.S.\ morphology
reweighted according to the CANGAROO-I acceptance and is shown at the
bottom left of Figure~\ref{fig_morphology_vela}, where the acceptance
distribution of H.E.S.S.\ \citep{aharonian06b} has not been considered
but is almost uniform within the field of view compared to the
CANGAROO-I acceptance. In this acceptance-reweighted morphology, the
head of the ``sea-horse'' shape, which corresponds to the CANGAROO-I
source as shown in the bottom right of
Figure~\ref{fig_morphology_vela}, is more emphasized than in the
original one. The CANGAROO-I morphology is obtained using the source
probability density function (PDF) method
\citep{yoshikoshi96}, in which a PDF for the source position with
respect to the image centroid, major axis, and \textit{asymmetry}
\citep{punch93, yoshikoshi96} is defined using Monte Carlo simulations
and a source distribution is made by adding up all PDFs of
gamma-ray-like events. The angular distance between the peak of the
acceptance-reweighted H.E.S.S.\ morphology and the CANGAROO-I position
is $0\fdg05$, which is comparable to the $1 \sigma$ error of the
CANGAROO-I position $\sim 0\fdg04$ \citep{yoshikoshi96}. Therefore,
the TeV gamma-ray signal detected by CANGAROO-I is possibly part of
HESS~J0835--455. However, the above acceptance-reweighted morphology
is still more extended than the CANGAROO-I source. This is possibly
due to the energy dependence of the gamma-ray morphology: the
threshold energy of the H.E.S.S.\ observations of Vela is 450\,GeV
which is an order of magnitude lower than that of CANGAROO-I (see
Section~\ref{sec_discussion_vela} for further discussion).

The integral flux from HESS~J0835--455 has been estimated using the
method described in Section~\ref{sec_flux}. This H.E.S.S.\ source is
diffuse extending up to the radius of $0\fdg8$ centered on
position~II ($\alpha = 8^\mathrm{h} 35^\mathrm{m} 00^\mathrm{s}$,
$\delta = -45\degr 36\arcmin$ (J2000)) defined by
\citet{aharonian06a}. The \textit{dis} and \textit{alpha} cuts are not
useful to estimate the flux of the same circular area as used by
H.E.S.S., and instead we must estimate the arrival direction of each
event as in the case of stereoscopic observations. Here, we estimate
the arrival direction as the most probable point of the source PDF
described above. The best gamma-ray selection criteria have been
obtained for the position offset by $0\fdg4$ from the camera center
since the position of the maximum emission of HESS~J0835--455 is
located around it on average. Simulation events have been generated
with a power-law spectrum of the index of $-1.45$ and an exponential
cutoff at 13.8\,TeV \citep{aharonian06a}. In the case of the 1993--1995
data, the numbers of \textsc{on}- and \textsc{off}-source
gamma-ray-like events falling into the circular area are 10,122 and
9621, respectively. The excess of 501 events corresponds to $3.6
\sigma$, which is less significant than that for the CANGAROO-I
position, owing to the inclusion of more background events in the
extended area. The acceptance $\epsilon$ depends on the source
distribution in the field of view, but first we simply assume that the
source is uniformly distributed in the circular area and $\epsilon$ is
averaged over it. The integral flux from HESS~J0835--455 ($0\fdg8$
radius centered on position~II) has thus been obtained as follows:
\ifapj
\begin{displaymath}
\begin{split}
F_\textrm{\scriptsize 93--95}(> 4.0 \pm 1.6\,\textrm{TeV}) = (3.28 \pm
0.92) \times 10^{-12}\\
\textrm{photons}\;\textrm{cm}^{-2}\,\textrm{s}^{-1}.
\end{split}
\end{displaymath}
\else
\begin{displaymath}
F_\textrm{\scriptsize 93--95}(> 4.0 \pm 1.6\,\textrm{TeV}) = (3.28 \pm
0.92) \times 10^{-12}\,
\textrm{photons}\;\textrm{cm}^{-2}\,\textrm{s}^{-1}.
\end{displaymath}
\fi
The stated error in the threshold energy is the systematic error, as
defined by \citet{muraishi00} except for the uncertainty of the
spectral index, whereas the error of the flux is only statistical. The
threshold energy of 4.0\,TeV is higher than the previous estimation of
2.5\,TeV \citep{yoshikoshi97}, owing to the harder spectrum used in
the simulations this time. For HESS~J0835--455, the two-dimensional
Gaussian profile of the diffuse emission has been given by
H.E.S.S. The average acceptance can more realistically be calculated
using this profile, as weights and the integral flux obtained using
this acceptance is
\ifapj
\begin{displaymath}
\begin{split}
F_\textrm{\scriptsize 93--95,G}(> 4.0 \pm 1.6\,\textrm{TeV}) = (3.05
\pm 0.86) \times 10^{-12}\\
\textrm{photons}\;\textrm{cm}^{-2}\,\textrm{s}^{-1}.
\end{split}
\end{displaymath}
\else
\begin{displaymath}
F_\textrm{\scriptsize 93--95,G}(> 4.0 \pm 1.6\,\textrm{TeV}) = (3.05
\pm 0.86) \times 10^{-12}\,
\textrm{photons}\;\textrm{cm}^{-2}\,\textrm{s}^{-1}.
\end{displaymath}
\fi
The difference between the above two fluxes is small and well within
the statistical errors.

\begin{figure}
\ifapj
\includegraphics[width=\linewidth]{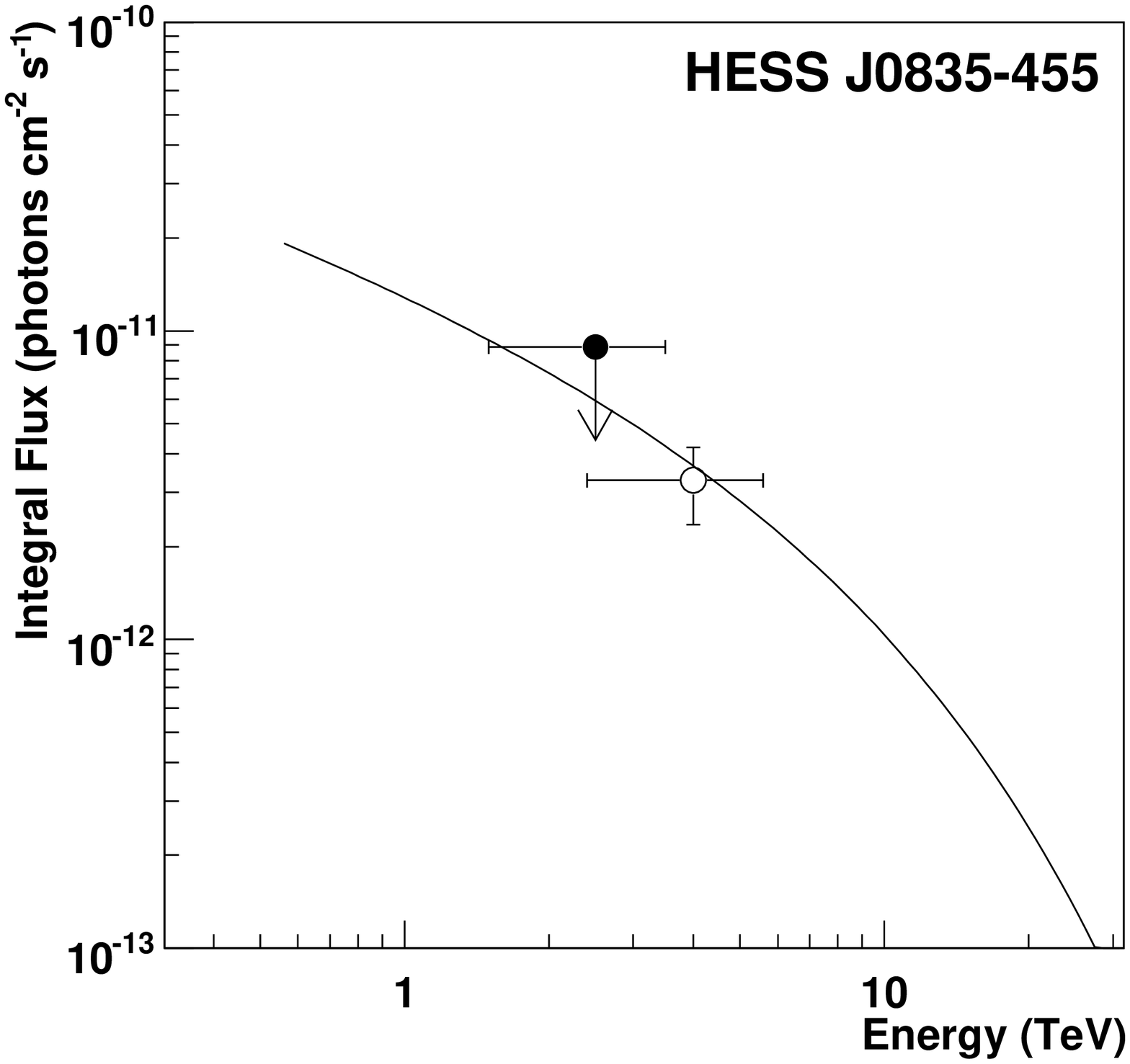}
\else
\plotone{f9.eps}
\fi
\caption{Integral gamma-ray fluxes from the Vela pulsar region
($0\fdg8$ radius centered on HESS~J0853--455) obtained from the
CANGAROO-I data. The open circle represents the integral flux from the
1993 to 1995 data, and the filled circle represents the 95\% CL upper
limit to the flux from the 1997 data. The horizontal bars represent
the systematic errors of the threshold energies. The solid line is the
integral spectrum converted from the differential spectrum of
HESS~J0853--455 measured by H.E.S.S.\ assuming a power law with an
exponential cutoff \citep{aharonian06a}. \label{fig_flux_vela}}
\end{figure}

The same calculation has been done also for the 1997 data: 2624
\textsc{on}-source and 2549 \textsc{off}-source events remained after
the selections and the excess of 75 events corresponds to $1.0
\sigma$. Since this excess is not statistically significant, the 95\%
CL upper limits to the integral flux have been calculated as follows:
\begin{displaymath}
F_\mathrm{97}(> 2.5 \pm 1.0\,\textrm{TeV}) < 8.87 \times
10^{-12}\,\textrm{photons}\;\textrm{cm}^{-2}\,\textrm{s}^{-1},
\end{displaymath}
\begin{displaymath}
F_\textrm{\scriptsize 97,G}(> 2.5 \pm 1.0\,\textrm{TeV}) < 8.33 \times
10^{-12}\,\textrm{photons}\;\textrm{cm}^{-2}\,\textrm{s}^{-1}.
\end{displaymath}
These fluxes are plotted in Figure~\ref{fig_flux_vela} as well as the
integrated H.E.S.S.\ spectrum \citep{aharonian06a}.
The estimated fluxes are all consistent with the H.E.S.S.\ spectrum
within statistical errors.

\section{Discussion} \label{sec_discussion}

The results of the reanalyses described in the previous section are
consistent with the recent H.E.S.S.\ results except for the
morphological difference in the Vela pulsar region, which is possibly
due to the difference of the threshold energy between the two
experiments. In other words, the previous CANGAROO-I results for
PSR~B1706--44 and SN~1006 claiming the significant detections of TeV
gamma rays have not been reproduced in the reanalyses. Therefore, we
have tried to apply the same analysis methods as used before to the
PSR~B1706--44 and SN~1006 data to determine why the old gamma-ray
signals were obtained, although not all details of the original
analysis are still available.

\subsection{Reproducibility of the Old PSR~B1706--44 Results}

Details of the analysis used to obtain the old PSR~B1706--44 results
are not given in the paper by \citet{kifune95}. Therefore, we have
used the analysis method described by \citet{tamura94} instead. The
known differences of the method from the standard analysis described
in Section~\ref{sec_analyses} are listed below.
\begin{enumerate}
\item Some noise cut levels are different: $\textit{size} \geq 100$,
$N_\mathrm{hit} \geq 4$, and $b$ (box noise parameter) $\leq 0.7$.
\item The gamma-ray selection criteria were defined as follows:
\begin{displaymath}
\begin{array}{rcccl}
        0\fdg01 & \leq & \textit{width}  & \leq & 0\fdg14 +
0\fdg045 \frac{\textit{size}}{1500} \\ 
         0\fdg1 & \leq & \textit{length} & \leq & 0\fdg6  \\
                &      & \textit{length} & \leq & 0\fdg2 + 0\fdg4
\frac{\textit{size}}{1500}          \\
0\fdg2          & \leq & \textit{dis}    & \leq & 0\fdg96 \\
\textit{length} & \leq & \textit{dis}    &      &         \\
                &      & \textit{alpha}  & \leq & 6\fdg75.
\end{array}
\end{displaymath}
\end{enumerate}
In the old analysis, the above conditions were applied only to the
PSR~B1706--44 data taken in 1993 August, and the details used in the
analysis of the other 1993 data are not given. Therefore, we applied
the above conditions only to the 1993 August data, but did not find
any significant \textit{alpha} peak around $0\degr$.

Subsequent examination of old internal collaboration
documents revealed the following facts about the 1993 August
analysis.
\begin{enumerate}
\item Only seven of the nine \textsc{on}-source runs taken in
1993 August were used in the analysis.
\item Calibrated positions of the tracking center
(Section~\ref{sec_tracking}) were not used since the calibration
method had not yet been established at that time. Instead, the
position of the tracking center was fixed in the camera at the
position of $(x,\ y) = (0\fdg1,\ -0\fdg125)$, which is offset from
the calibrated tracking centers by $0\fdg13$ on average.
\item In the image cleaning, the threshold for ADC values was fixed at 
5 ADC counts.
\item Noise cut levels different from those in the standard analysis
were used: $100 \leq \textit{size} < 5000$, $N_\mathrm{hit} \geq 4$,
and $b \leq 0.7$. Also, the effect of the ADC offset noise (see
Section~\ref{sec_image_cleaning}) was not corrected.
\item Another noise rejection called the ``region cut'' was
applied. Events with centroids located within the areas of ``noisy''
box units were rejected in this cut. The ``noisy'' boxes were
determined run by run and the gross number of the ``noisy'' boxes was
eight in 1993 August.
\item The gamma-ray selection criteria were defined as follows:
\begin{displaymath}
\begin{array}{rcccl}
        0\fdg01 & \leq & \textit{width}  & \leq & 0\fdg25 \\
                &      & \textit{width}  & \leq & 0\fdg1 + 0\fdg03
\frac{\textit{size}}{1500} \\ 
         0\fdg1 & \leq & \textit{length} & \leq & 0\fdg6  \\
                &      & \textit{length} & \leq & 0\fdg2 + 0\fdg45
\frac{\textit{size}}{1500} \\ 
         0\fdg2 & \leq & \textit{dis}    & \leq & 0\fdg96 \\
\textit{length} & \leq & \textit{dis}.   &      &
\end{array}
\end{displaymath}
The selection levels were modified slightly run by run, but no further
information beyond this is available.
\end{enumerate}
One can note that some of the above selection levels are different to those
of \citet{tamura94}. There seems to be some small changes in the
analysis between the papers of \citet{tamura94} and \citet{kifune95}.

\begin{figure}
\ifapj
\includegraphics[width=\linewidth]{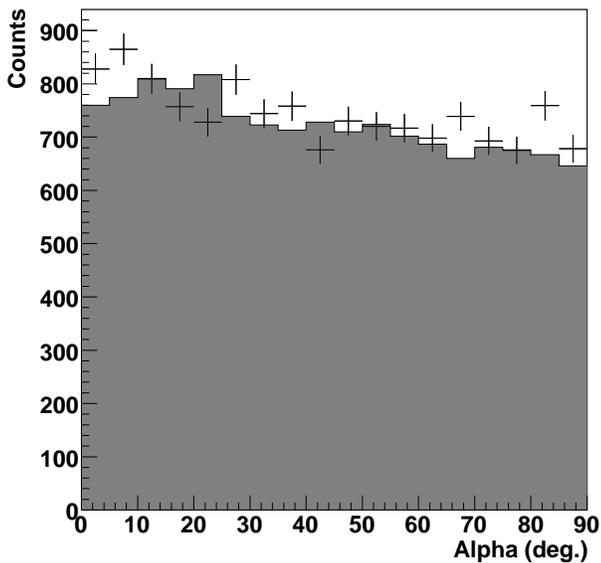}
\else
\plotone{f10.eps}
\fi
\caption{Distributions of the orientation parameter \textit{alpha} for
the PSR~B1706--44 data taken in 1993 August. The analysis method is
based on the descriptions in the old internal documents to try to
reproduce the result obtained by \citet{kifune95}. The histogram with
the error bars and the gray solid histogram are the \textsc{on}- and
\textsc{off}-source distributions, respectively. The number of events
remaining after selections is similar to that in Figure~1 (d) of
\citet{kifune95}, although the bin sizes of the histograms
differ. \label{fig_alpha_1706_old}}
\end{figure}

We have applied the above conditions to the 1993 August data and the
obtained \textit{alpha} distributions are shown in
Figure~\ref{fig_alpha_1706_old}.
Some excess of the \textsc{on}-source events has been found near
$\textit{alpha} = 0\degr$ and its statistical significance is $2.8
\sigma$ taking the events of \textit{alpha} smaller than
$10\degr$. However, the excess is not as sharp as the old
result. Regarding the significance, \citet{kifune95} used a different
definition, in which the background level was estimated from the
\textsc{on}-source events of $\textit{alpha} > 30\degr$ since their
\textit{alpha} distributions appeared to be flat. Following this
definition, the significance is enhanced by a factor of $\sqrt{2 / (1
+ \alpha)} \sim 1.3$, where $\alpha = 1 / 6$ is the normalization
factor. It is also found from Figure~1 of \citet{kifune95} that the
amount of the 1993 August data used in our reanalysis is roughly $1 /
5$ of the data used in the old analysis. If we simply scale the total
significance using the above factors, it turns out to be $2.8 \sigma
\times 1.3 \times \sqrt{5} \sim 8 \sigma$, which is not so different
from the original claim of $12 \sigma$ despite the very rough
estimation.  However, we think that this excess is not due to gamma
rays from PSR~B1706--44 since the tracking center $(0\fdg1,\
-0\fdg125)$ in the camera is not very close to the calibrated
tracking centers and the Cherenkov images have been left affected by
the ADC offset noise. Only a broad \textit{alpha} peak smeared by the
source offsets and the ADC noise would be expected if it were really
due to gamma rays.

A possible reason for the apparently significant gamma-ray-like
signal in the previous analysis is that an initial excess has been
enhanced by the complicated selection levels, and/or their run-by-run
modifications but the consequent statistical penalty for the
number of degrees of freedom used has been neglected.
For example, the ``region cut'' was introduced in the name of noise
rejection but the ``noisy'' boxes were not selected independently of
the gamma-ray-like signal. If the $8 \times 0.4 \sim 3$ ``noisy''
boxes are randomly selected among the total number of boxes, 28
boxes per camera $\times$ 7 runs $\times$ 0.4 $\sim 78$, the number of
possible combinations in the ``region cut'' is ${78 \choose 3} =
76,076$, where the factor 0.4 is the ratio of the effective camera area
after the \textit{dis} selection of $\textit{dis} \leq 0\fdg96$. The
total gamma-ray-like signal (number of gamma-ray-like excess events)
$z$ of a combination is represented as
\begin{equation}
z = \sum_{i = 1}^{78} x_i - \sum_{j = 1}^{3} y_j,
\end{equation}
where $x_i \in X = \{x_1,\ x_2,\ \dots,\ x_{78}\}$ is the signal of
the $i$th box (number of excess events of which the centroids are
located within the box area) and $y_j \in X$ is also a signal of a box
but the index is renumbered for rejected boxes. Assuming that $x_i$ is
normally distributed with mean 0 and standard deviation $\sigma$, any
$z$ is also normally distributed with mean 0 and standard deviation
$\sigma' = \sigma \sqrt{78 - 3} = \sqrt{75} \sigma$. However, if the
maximum $z$ is chosen among all combinations, $z_\mathrm{max}$ is no
longer distributed around 0, and in this case, the distribution has a
positive offset of about $1 \sigma'$. These kinds of offsets are
accumulated by the number of degrees of freedom, and the signal can be
made apparently significant just from the background
fluctuations. Therefore, we suspect that the very complicated
conditions in the old PSR~B1706--44 analysis described above could
supply sufficient degrees of freedom to generate the reported signal.

We have also investigated how much the significance can be enhanced in
the real data by varying the region cut. The selection criteria used
in the old analysis have been adopted except the region cut, and the
number of rejected boxes in varied region cuts has been fixed at 8.
In the real application, the above approximation using the factor 0.4
is not possible because the region cut is discrete, and the total
number of boxes increases to 147, which is derived from 21 boxes per
camera fully or partially contained in the \textit{dis} selection
area, multiplied by the 7 runs. The \textit{dis} selection is the same
in both, but they differ in how to count boxes partially contained in
the \textit{dis} selection area. The number of boxes per camera before
region cuts is $28 \times 0.4 = 11.2$ in the toy model, and 4 (fully
contained) $+$ 17 (partially contained) $= 21$ in the real
application. A total of $10^6$ combinations has randomly been sampled
from the all ${147 \choose 8} \sim 4.5 \times 10^{12}$ combinations
and their significances have been calculated after applying the
corresponding region cuts. They are normally distributed around a mean
significance of $2.4 \sigma$ with a root mean square of $0.24
\sigma$. The maximum significance is $3.5\sigma$ with this number of
samples. The significance obtained without using the region cut is
$2.5 \sigma$ and thus the increase of the significance with the region
cut actually used in the old analysis is $0.3 \sigma$ of the total
$2.8 \sigma$, which is in fact a positive enhancement, although
smaller than the possible maximum offset estimated above. However, the
probability of obtaining a significance greater than $2.8 \sigma$ with
any region cut of 8 boxes is 6.4\%, which is small to be interpreted
by chance.

\subsection{Reproducibility of the Old SN~1006 Results}
\label{sec_old_1006}

Details of the old SN~1006 analysis are not given in full in
\citet{tanimori98}, and so we have followed the analysis method
described by \citet{kamei98} instead. The gamma-ray selection
criteria were as follows:
\begin{enumerate}
\item for the 1996 April data,
\begin{displaymath}
\begin{array}{rcccl}
        0\fdg04 & \leq & \textit{width}  & \leq & 0\fdg17 \\
        0\fdg10 & \leq & \textit{length} & \leq & 0\fdg48 \\
           0.64 & \leq & \textit{conc}   &      &         \\
         0\fdg1 & \leq & \textit{dis}    &      &         \\
\textit{length} & \leq & \textit{dis}    &      &         \\
                &      & \textit{alpha}  & \leq & 15\degr,
\end{array}
\end{displaymath}
\item for the 1996 June data,
\begin{displaymath}
\begin{array}{rcccl}
                     0\fdg04 & \leq & \textit{width}  & \leq & 0\fdg17 \\
                     0\fdg10 & \leq & \textit{length} & \leq & 0\fdg38 \\
                      0\fdg4 & \leq & \textit{dis}    & \leq & 1\fdg2  \\
3.3 \textit{length} - 1\fdg0 & \leq & \textit{dis}    &      &         \\
                             &      & \textit{alpha}  & \leq & 15\degr,
\end{array}
\end{displaymath}
\item for the 1997 data and events of $\textit{size} \leq 2000$,
\begin{displaymath}
\begin{array}{rcccl}
                              0\fdg04 & \leq & \textit{width}  &
\leq & 0\fdg13                               \\
-0\fdg23 + 0\fdg13 \log \textit{size} & \leq & \textit{length} &
\leq & 0\fdg07 + 0\fdg084 \log \textit{size} \\
              0.23 \log \textit{size} & \leq & \textit{conc}   &
\leq & 0.42 + 0.14 \log \textit{size}        \\ 
                               0\fdg5 & \leq & \textit{dis}    &
\leq & 1\fdg1                                \\ 
                      \textit{length} & \leq & \textit{dis}    &
     &                                       \\ 
                                      &      & \textit{alpha}  &
\leq & 15\degr,
\end{array}
\end{displaymath}
\item for the 1997 data and events of $\textit{size} > 2000$,
\begin{displaymath}
\begin{array}{rcccl}
        0\fdg04 & \leq & \textit{width}  & \leq & 0\fdg13 \\
         0\fdg2 & \leq & \textit{length} & \leq & 0\fdg07 + 0\fdg084
\log \textit{size} \\
           0.72 & \leq & \textit{conc}   & \leq & 0.42 + 0.14 \log
\textit{size}      \\
         0\fdg5 & \leq & \textit{dis}    & \leq & 1\fdg1  \\ 
\textit{length} & \leq & \textit{dis}    &      &         \\ 
                &      & \textit{alpha}  & \leq & 15\degr,
\end{array}
\end{displaymath}
\end{enumerate}
where the definition of \textit{conc} is different from that
originally used by \citet{weekes89} and is the fraction of the light
detected by the camera that is contained in the brightest half of the
triggered pixels. The other conditions different from the standard
analysis described in Section~\ref{sec_analyses} are listed below.
\begin{enumerate}
\item The \textsc{on}- and \textsc{off}-source data were not matched
in terms of the geometrical coordinates (azimuth and elevation). More
data of the longer observation times listed in Table~\ref{tab_data}
without brackets were used. The background levels of the
\textsc{on}-source \textit{alpha} distributions were estimated from
the \textsc{off}-source distributions normalized by the observation
times or from the extrapolation of the \textsc{on}-source
distributions of $\textit{alpha} > 30\degr$.
\item In the image cleaning, the threshold for ADC values was fixed to
be 15 ADC counts.
\item The box noise cut was not used. Box noise events were manually
removed looking at image centroid and shower rate distributions
instead, but no details of rejected areas or periods remain. The ADC
offset correction was not done. The definition of the camera edge cut
was different and not based on the image distance from the camera
center. Instead, images for which the centroid was located in the area
of the outermost pixels were rejected.
\item The discharge noise cut introduced in Section~\ref{sec_1006}
was not used in the old analysis but instead events with
centroids located in the bottom quarter of the camera were removed
since most of the noisy pixels were located there. This cut was
applied only to the 1996 data.
\item The scaler cut to eliminate the bright star effect was more
complicated than described in Section~\ref{sec_1006}. The mean and
standard deviation of scaler values with no bright star in the field
of the pixel were first calculated for each pixel and each run. Then
signals with scaler values greater than $8 \sigma$ over the mean 
value were rejected from the images when the distances of the pixel
from bright stars were smaller than $0\fdg3$.
\end{enumerate}
In the case of the old SN~1006 analysis, calibrated positions of the
tracking center had been used and therefore the accuracy of the object
position was better than in the case of the old PSR~B1706--44 analysis.

\begin{figure}
\ifapj
\includegraphics[width=\linewidth]{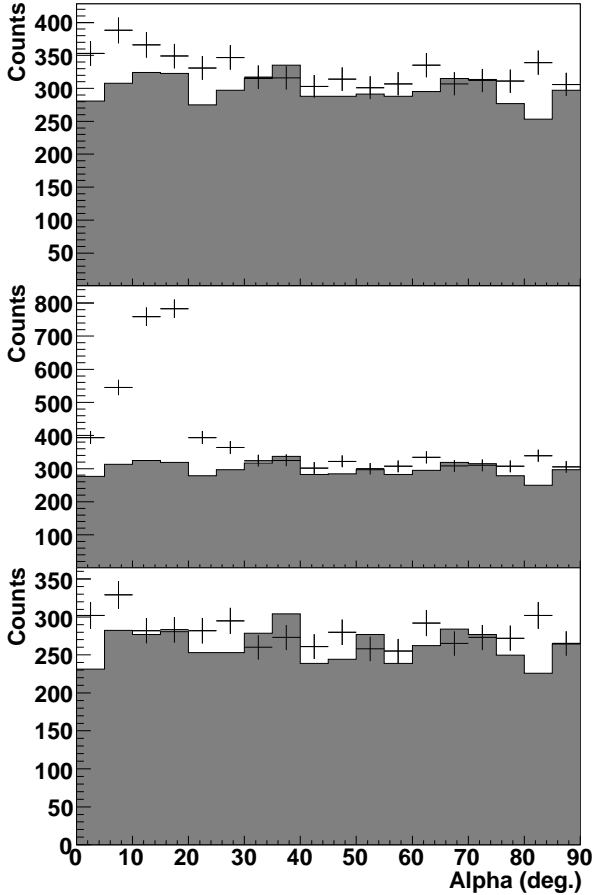}
\else
\epsscale{0.5}
\plotone{f11.eps}
\epsscale{1}
\fi
\caption{Distributions of the orientation parameter \textit{alpha} for
the SN~1006 data taken in 1996 obtained using the old analysis
method. The \textit{alpha} values have been calculated with respect to
the NE rim of the SNR. The \textsc{on}- and \textsc{off}-source
distributions are represented by the histograms with the error bars
and the gray solid histograms, respectively. Top: distributions
obtained using the old analysis method described by \citet{kamei98},
middle: distributions obtained using the same method but without the
$8 \sigma$ scaler cut, bottom: distributions obtained applying the
discharge noise cut to the middle
distributions. \label{fig_alpha_1006_old}}
\end{figure}

The above analysis procedure has been applied to the 1996 data, but the
box noise cut has been used instead of the manual rejection of the
noise that is impossible to reproduce. The obtained \textit{alpha}
distributions with respect to the NE rim are shown at the top of
Figure~\ref{fig_alpha_1006_old}, where the distribution of
\textsc{off}-source has been normalized to that of \textsc{on}-source
using the observation time.
A broad excess of the \textsc{on}-source events is found near
$\textit{alpha} = 0\degr$ with the statistical significance of $3.6
\sigma$ below $\textit{alpha} = 15\degr$. The significance increases
to $4.7 \sigma$ with the different assumption of the background level
used by \citet{tanimori98}, in which the background level was
estimated from the ``flat'' region of the \textit{alpha} plot
($\textit{alpha} > 30\degr$) of the \textsc{on}-source data. This
result is similar to the old result obtained by \citet{tanimori98} in
terms of the remaining number of events, the statistical significance
level, and the \textit{alpha} peak broadness. The middle of
Figure~\ref{fig_alpha_1006_old} is the
\textit{alpha} distributions of the same data obtained with the same
method except for the $8 \sigma$ scaler cut. The remarkable excess
around $\textit{alpha} = 15\degr$ in this figure has however
disappeared further applying the discharge noise cut described in
Section~\ref{sec_1006} as shown in the bottom of
Figure~\ref{fig_alpha_1006_old}. The $8 \sigma$ scaler cut was
originally introduced to reduce the effect of the bright stars in the
field of view, but also has the effect of reducing the discharge noise
since noisy pixels also have larger scaler values.  Therefore, it is
possible that the excess of the top figure consists of discharge noise
events remaining after the $8 \sigma$ scaler cut, which is moderately
loose for discharge noise.

\begin{figure}
\ifapj
\includegraphics[width=\linewidth]{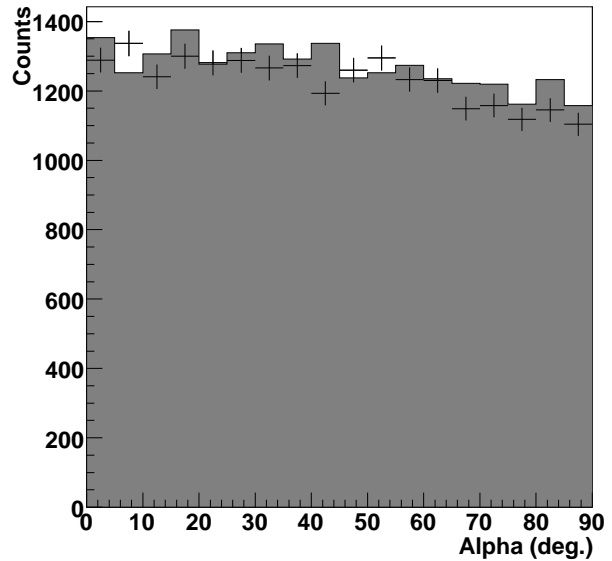}
\else
\plotone{f12.eps}
\fi
\caption{Distributions of the orientation parameter \textit{alpha} for
the SN~1006 data taken in 1997 obtained using the old analysis
method. The \textit{alpha} values have been calculated with respect to
the NE rim of the SNR. The \textsc{on}- and \textsc{off}-source
distributions are represented by the histogram with the error bars and
the gray solid histogram,
respectively. \label{fig_alpha_1006_1997_old}}
\end{figure}

The above old analysis method has also been applied to the 1997 data,
and Figure~\ref{fig_alpha_1006_1997_old} shows the 
\textit{alpha} distributions obtained.
In contrast to the 1996 result, this 1997 result is quite different
from the old result. No gamma-ray-like excess has been found this time
and the overall features of the old result have not been reproduced. The
number of remaining background events are about two times more than
that of the old 1997 \textit{alpha} plot, and therefore some 
additional selection criteria must have been used in the old
analysis. However, no further information about the old analysis is
available.

\subsection{Vela} \label{sec_discussion_vela}

It is notable that all of the significances estimated in this paper
have decreased from those previously reported. There are two possible
reasons for this: the previous signal for the Vela pulsar region had
also been enhanced with trials in the parameter domain and/or
gamma-ray events have been overcut in the present analysis since the
gamma-ray selection criteria used this time are possibly too tight
owing to the unmodeled effects in the simulations (see the
Appendix~\ref{sec_simulation} for more discussion). The signal from
the CANGAROO-I position near the Vela pulsar obtained using the 1997
data is enhanced from $2.5 \sigma$ with the standard analysis in this
paper to $\sim 4 \sigma$ previously reported by \citet{yoshikoshi98},
if the \textit{size} and \textit{length} selection levels are looser
than their standard values ($\textit{size} \geq 180$ ADC counts and
$\textit{length} \leq 0\fdg45$ were used by \citet{yoshikoshi98})
and more data before matching \textsc{on}- and \textsc{off}-source
observations (listed without brackets in Table~\ref{tab_data}) are
used. We have had no standard candle in the southern hemisphere strong
enough compared to the sensitivity of the 3.8\,m telescope and cannot
fine-tune our simulation code using a real gamma-ray signal. It is
consequently difficult to distinguish between these two reasons.
However, in the case of the Vela 1997 data, $2.5 \sigma$ is thought to
be a more conservative and reliable estimate than the previous value
since it has been obtained with a common analysis method and by
reducing unaccounted for degrees of freedom. The optimum level of the
\textit{length} selection is sensitive to the source location in the
camera as seen in Figure~\ref{fig_best_q}, and thus the
\textit{length} selection level of $0\fdg45$ used by
\citet{yoshikoshi98} was conservatively selected considering the
search area of the $2\degr \times 2\degr$ field of view. However, the
\textit{length} selection level of $0\fdg45$ is very different from
the optimized level of $\sim 0\fdg3$ for the CANGAROO-I position,
and too loose to justify the above signal enhancement if we assume our
simulations to be acceptable for the optimizations of the selection
levels.

\citet{dazeley01b} adopted a similar analysis method in
which they have optimized the parameter selection criteria so as to
maximize the quality factor using their simulations. However, they
have obtained a null result applying their method to the Vela 1997
data. This difference is probably due to the different input
conditions in their simulations which are described in the
Appendix~\ref{sec_simulation} in detail. \citet{dazeley01b} have
estimated the $3 \sigma$ upper limit to the integral flux from the
Vela pulsar region to be $2.5 \times
10^{-12}\,\textrm{photons}\;\textrm{cm}^{-2}\,\textrm{s}^{-1}$ above
2.7\,TeV, which is about a factor of two smaller than the integral
flux from HESS~J0835--455 obtained by H.E.S.S.\ above the same
threshold energy. However, their upper limit is not inconsistent with
the H.E.S.S.\ result since only part of the emission region of
HESS~J0835--455 was included in their estimation as in the case of the
CANGAROO-I result.

To investigate the above consistency more quantitatively, we have
compared the integral fluxes between CANGAROO-I and H.E.S.S., changing
the integration region around the CANGAROO-I
position. Figure~\ref{fig_accumulated_flux_vela} shows the integral
fluxes from the Vela pulsar region as a function of the radius of the
integration region.
\ifapj
\begin{figure*}
\includegraphics[width=\linewidth]{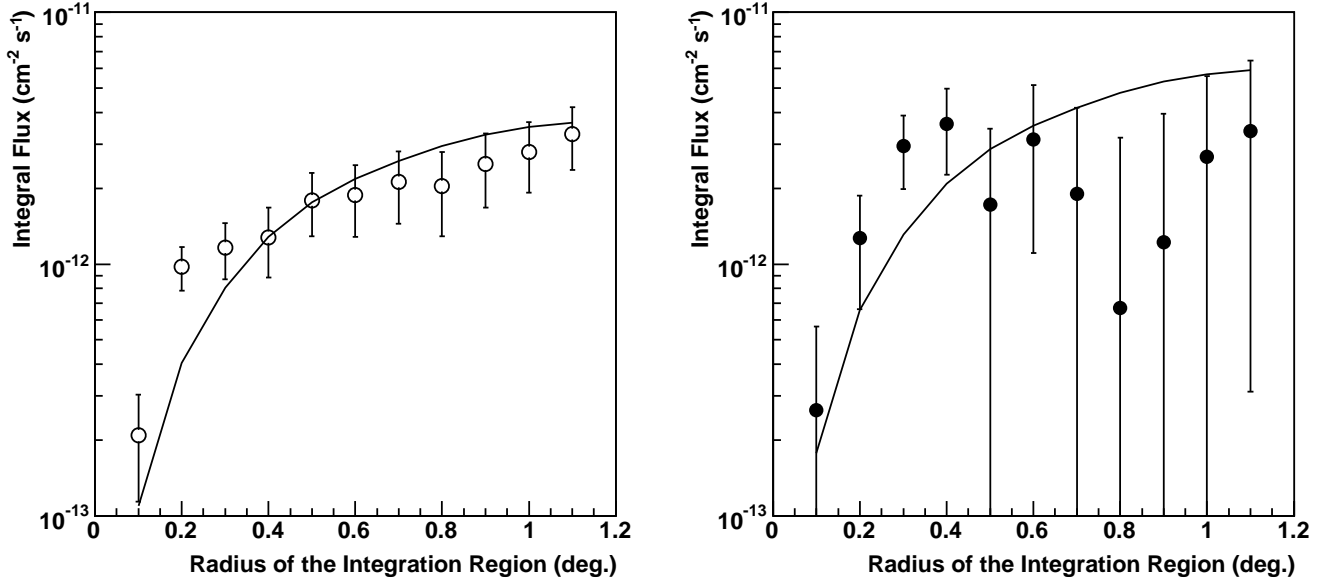}
\else
\begin{figure}
\plotone{f13.eps}
\fi
\caption{Integral gamma-ray fluxes from the Vela pulsar region as a
function of the radius of the integration region. The circular
integration regions have been centered on the CANGAROO-I position and
limited inside of the H.E.S.S.\ integration region shown in
Figure~\ref{fig_morphology_vela}. The left and right plots are the
results of the 1993--1995 data ($E_\mathrm{th} = 4.0$\,TeV) and the 1997
data ($E_\mathrm{th} = 2.5$\,TeV), respectively. The CANGAROO-I results
are represented by the markers with the error bars, while the
H.E.S.S.\ results are shown by the
lines. \label{fig_accumulated_flux_vela}}
\ifapj
\end{figure*}
\else
\end{figure}
\fi
The circular integration regions have been centered on the CANGAROO-I
position and limited to the inside of the H.E.S.S.\ integration region
shown in Figure~\ref{fig_morphology_vela}. The CANGAROO-I fluxes have
been calculated considering the acceptance distribution in the field
of view. The source distribution has been assumed to be uniform in
estimating the average acceptance in the integration region, and as
shown in Section~\ref{sec_vela}, the fluxes are reduced by about 10\%
if a Gaussian source distribution is used instead. The H.E.S.S.\
fluxes have simply been obtained using the following formula:
\begin{equation}
F_r(> E_\mathrm{th}) = \frac{N_r}{N} F(> E_\mathrm{th}),
\end{equation}
where $F(> E_\mathrm{th})$ and $N$ are the integral flux and excess
counts for HESS~J0835--455, respectively, and $F_r(> E_\mathrm{th})$
and $N_r$ are those integrated only inside a radius $r$ from the
CANGAROO-I position. One can note that the CANGAROO-I fluxes exceed
the H.E.S.S.\ fluxes in the proximity of the CANGAROO-I position (the
most significant excess of $3.0 \sigma$ is found at the radius of
$0\fdg2$ from the 1993--1995 data) but the fluxes are statistically
consistent with each other at large $r$ in the both data sets, i.e.,
the profile of the CANGAROO-I source is more compact than that of the
H.E.S.S.\ source. This tendency is acceptable if the gamma-ray
morphology is energy dependent, since the threshold energies of
CANGAROO-I are 5--10 times higher than that of H.E.S.S. There is a
known example, the pulsar wind nebula HESS~J1825--137, in which the
morphology is energy dependent and its higher energy emission is more
compact and closer to the pulsar position
\citep{aharonian06c}.

\section{Conclusions} \label{sec_conclusions}

We have reanalyzed the CANGAROO-I data of PSR~B1706--44, SN~1006, and
Vela using a consistent analysis method, in which the gamma-ray
selection criteria have been optimized using gamma-ray simulations and
\textsc{off}-source data. The analysis method is the simplest using a
single cut or two for each image parameter and almost free from
arbitrary degrees of freedom. The previously reported signals from
PSR~B1706--44 and SN~1006 have not been reproduced with this analysis
method using the same data, whereas the signal from the CANGAROO-I
position offset by $0\fdg13$ to the southeast from the Vela pulsar
has still been detected at the $4.5 \sigma$ level. The new upper
limits to the integral fluxes from PSR~B1706--44 and SN~1006 at the
95\% CL are higher than the upper limits obtained by H.E.S.S.\ and so
there is formally no inconsistency between them. The emission profile
of the signal from the Vela pulsar region obtained from the CANGAROO-I
data has been compared with that of H.E.S.S.\ considering the
acceptance in the field of view of the 3.8\,m telescope, which quickly
drops toward the edge of the imaging camera. The expected emission
profile for CANGAROO-I obtained by reweighting the diffuse emission
from HESS~J0835--455 with the CANGAROO-I acceptance distribution in
the field of view is similar to the observed CANGAROO-I sky map, but
the latter profile seems to be more compact. This difference in shape
is possibly due to the fact that the threshold energy of the
CANGAROO-I observations was an order of magnitude higher than that of
H.E.S.S.\ and the energy dependent morphology of HESS~J0835--455
should be studied in more detail.

We have attempted to reproduce the previous CANGAROO-I analyses for
PSR~B1706--44 and SN~1006 using the documented analysis methods to
investigate why the significant gamma-ray signals had been
obtained. ``Gamma-ray-like'' excesses similar to the old results have
been found with the analyses carried out for the 1993 August data of
PSR~B1706--44 and the 1996 data of SN~1006, although they are still
less significant than the old results. The old results of the other
data have not been reproduced on the basis of available information,
but we can conclude the following.
\begin{enumerate}
\item The previous reports are inconsistent with the results obtained 
here and as some important calibration results such as the effect of
the ADC offset noise were not considered in the former analyses, their
significance must be questioned.
\item The previous analyses were unduly complicated, resulting in a
large number of degrees of freedom in the background event
reduction. Changing the analysis conditions depending on the data sets
added to the degrees of freedom, but no reasons for the differences
were clarified independently of the \textsc{on}-source data. Selecting
noise rejection and/or gamma-ray selection levels depending on the
\textsc{on}-source data itself may have resulted in the apparently
significant signals.
\end{enumerate}

\acknowledgments

We thank all members of the CANGAROO-I team for their efforts in
building and maintaining the telescopes and helping take the data used
in these analyses. We also thank Dr.\ Bruno Kh\'{e}lifi, Dr.\ Conor
Masterson, Dr.\ Gavin Rowell, and the H.E.S.S.\ Collaboration for
providing us with the sky map of HESS~J0835--455, the upper limit data
of PSR~B1706--44 and SN~1006, and related information. Thanks
to the referee for useful comments helping clarify the paper. This
work has been supported by a Grant-in-Aid for Scientific Research of
the Ministry for Education, Culture, Sports, Science, and Technology
(Japan) and the Australian Research Council.

\appendix

\section{Monte Carlo Simulations} \label{sec_simulation}

The Monte Carlo simulation code consists of the following three parts:
a) generation of EASs, b) Cherenkov light emission from the EASs and
its attenuation in the atmosphere, and c) response of the telescope
system. To generate EASs, GEANT 3.21 \citep{cern93} has been used with
the target atmosphere modeled by 80 homogeneous layers, in which the
densities have been calculated at the average heights using the US
standard atmosphere. Primary gamma rays and protons have been injected
at the height 50\,km a.s.l., under which the effect of the
homogeneous geomagnetic field of 57.7\,$\mu$T that has been calculated
for Woomera using the International Geomagnetic Reference Field (IGRF)
model has been incorporated. Cherenkov photons are approximately
emitted from the end point of each charged particle path only if they
are falling into the telescope mirror area in order to reduce the
computation time. Rayleigh and Mie scatterings have been considered as
the extinction of Cherenkov photons in the atmosphere. The mean free
path of the Rayleigh scattering is
$2974\,\textrm{g}\;\textrm{cm}^{-2}$ at 400\,nm, and the scale height
and mean free path of the Mie scattering have been assumed to be
1.2\,km and 14\,km, respectively \citep{baltrusaitis85}. Scattered
photons have been neglected, i.e., deemed to be attenuated. The
computation time is further shortened by virtually reducing the number
of generated photons multiplied by the following factors in advance:
\begin{equation}
N' = N \epsilon_t \epsilon_\phi \epsilon_r \epsilon_q,
\end{equation}
where $N$ and $N'$ are the numbers of Cherenkov photons before and
after the reduction, respectively, and $\epsilon_t$, $\epsilon_\phi$,
$\epsilon_r$, and $\epsilon_q$ are the atmospheric transmission factor
due to the above scatterings, the fraction of the azimuthal angle
range of the telescope viewed by the charged particle, the mirror
reflectivity, and the quantum efficiency of the photocathode of the
PMTs ($\sim 20$\%), respectively.

The telescope optics has been modeled on a parabolic reflector causing
the coma aberration on the focal plane plus an on-axis Gaussian blur.
The blur spot size was measured using a CCD camera, with which an
image of an on-axis bright star was taken and approximated to the
two-dimensional Gaussian of $\sigma = 0\fdg04$. Photoelectron signals
due to Cherenkov radiation and incident on each PMT photocathode are
integrated together with the random NSB signals and converted to a
electronic waveform assuming that the signals have a common triangular
pulse shape and considering their arrival time differences. Then,
event triggers are examined with the conditions of the discrimination
levels of three photoelectrons for each pixel and five simultaneous hit
pixels. The former discrimination level has been calibrated comparing
scaler values (discrimination rate per 1\,ms) exposed to the NSB light
with the standard NSB brightness $2.55 \times
10^{-4}\,\textrm{erg}\;\textrm{s}^{-1}\,\textrm{cm}^{-2}\,\textrm{sr}^{-1}$
($430 \sim 550$\,nm) compiled by \citet{jelley58}. The latter for the
pixel multiplicity has been determined from lower cutoffs of
$N_\mathrm{hit}$ distributions as well as its fluctuation of $\pm 1$
pixel. One photoelectron signal corresponds to $5.0 \pm 0.7$ ADC
counts, which has been determined by equating lower cutoffs of ADC
distributions with the discrimination level of three photoelectrons. This
conversion factor is consistent with that calculated from parameters
of the electronics. Finally, ADC and TDC values of individual pixels
are calculated using the 50\,ns ADC gate width and the 0.24\,ns TDC
resolution, and recorded in the same format as observed data.

\begin{figure}
\ifapj
\includegraphics[width=\linewidth]{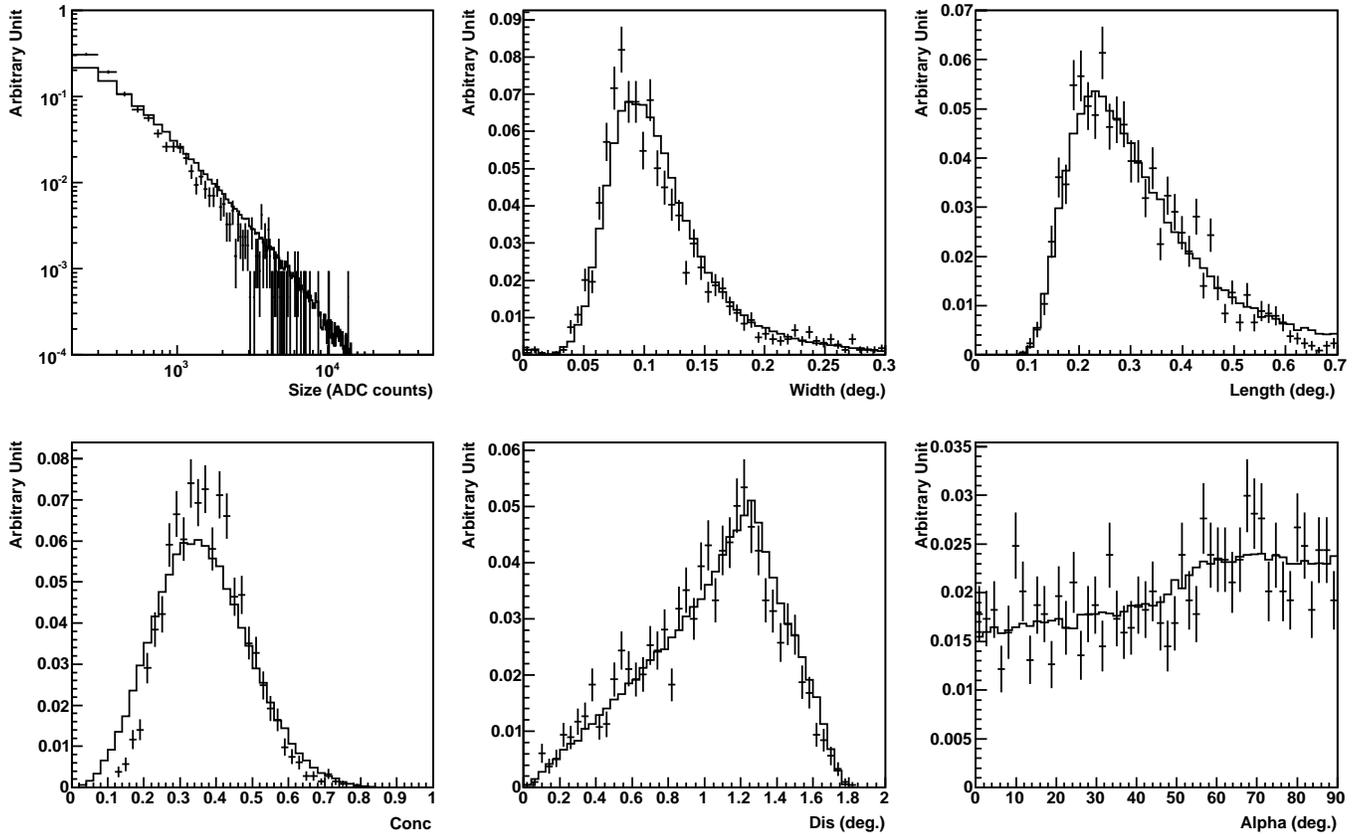}
\else
\plotone{f14.eps}
\fi
\caption{Distributions of the image parameters \textit{size} (top
left), \textit{width} (top middle), \textit{length} (top right),
\textit{conc} (bottom left), \textit{dis} (bottom middle), and
\textit{alpha} (bottom right) of background hadronic events. The
simulated and observed distributions are represented by the histograms
with the error bars and the solid histograms, respectively.  The
simulated histograms have been made injecting $10^6$ protons with the
power-law energy spectrum of the index $-2.7$, while the observed
histograms have been obtained using the \textsc{off}-source data of
Vela taken in 1997. Simple noise rejections, $b \leq 0.8$,
$\textit{size} \geq 200$ ADC counts, and $N_\mathrm{hit} \geq 5$, have
been applied before filling the histograms, which are all normalized
to 1 after filling. \label{fig_sim_obs_param}}
\end{figure}

To check the reliability of the simulations, we have tried to
reproduce the trigger rate and image parameter distributions of
background events. The trigger rate $R$ can be calculated using the
following formula:
\begin{equation}
R = A_0 \Omega_0 \epsilon \int_{E_\mathrm{min}}^{E_\mathrm{max}} f(E)
dE,
\end{equation}
where $f$ is the differential cosmic ray flux, which is integrated
over the simulated energy range between $E_\mathrm{min} = 500$\,GeV
and $E_\mathrm{max} = 100$\,TeV, $A_0 = \pi r_\mathrm{max}^2 \cos
\theta$ is the input area of the maximum radius $r_\mathrm{max} =
300$\,m on the ground with the input zenith angle $\theta$, $\Omega_0
= 2 \pi (1 - \cos
\Theta_\mathrm{max})$ is the input solid angle of the maximum offset
angle $\Theta_\mathrm{max} = 3\degr$ from the pointing direction, and
$\epsilon = N_\mathrm{triggered} / N_\mathrm{input}$ is the trigger
efficiency. The cosmic ray all particle flux has been obtained to be
$f(E) = 3.93 \times 10^4 (E /
1\,\textrm{GeV})^{-2.73}\,\textrm{m}^{-2}\,\textrm{s}^{-1}\,\textrm{sr}^{-1}\,\textrm{GeV}^{-1}$
by fitting a power-law function to the direct observation data
measured by \citet{ichimura93} between 500\,GeV and 500\,TeV. The
input primary particles are only protons, and so it is possible that
the simulated trigger rate is somewhat overestimated in comparison to
the observed one. The mirror reflectivity and the average zenith angle
used here are 75\% and $\theta = 15\degr$, respectively, to compare
the result with the Vela 1997 observations. A total of 3775 events
resulted in triggers out of 10$^6$ inputs and $\epsilon = 3775 / 10^6
= (3.78 \pm 0.06) \times 10^{-3}$, where the error is statistical
only. Thus, the expected trigger rate is estimated to be $4.32 \pm
0.07$\,Hz, which is somewhat higher than the real trigger rate of
2--3\,Hz. However, this result is not inconsistent with observations
since the fraction of protons in the cosmic ray all particle flux is
about 40\% \citep{ichimura93} and the trigger efficiencies for heavier
particles are much smaller than that for protons
\citep{aharonian99}. Image parameter distributions of the above
simulated data are shown in Figure~\ref{fig_sim_obs_param} in
comparison with those of the Vela 1997 data.
The observed distributions are reasonably reproduced by the simulations
except for the \textit{size} and \textit{conc} distributions. The same
tendency in \textit{conc} has also been shown by \citet{dazeley01a},
and they have pointed out the possibility that the difference in
\textit{conc} is due to electronic cross-talk.

\citet{dazeley01a} also tuned their simulation code by matching
trigger rates and image parameter distributions between simulations
and observations. However, they needed to introduce a signal loss
factor to do this, which is a free parameter based on the assumption
that signal intensities have been lost in the electronics. Using such
a factor is inconsistent with the above-mentioned calibration results,
and our simulations reproduce the observed features without
it. Low-energy events and the responses to them predominantly
determine trigger rates and image parameter distributions since the
spectrum of primary cosmic rays has a steep power-law form and
compensating for the disagreement at low energies with a single
scaling factor possibly results in distortion at higher energies
instead. Note that the cosmic ray spectrum can be reproduced at
relatively high energies using CANGAROO-I data without the loss factor
as shown by \citet{yoshikoshi99}. The differences between our
simulations and those of \citet{dazeley01a} are summarized in
Table~\ref{tab_simulation_dp}. A major difference other than the
signal loss factor is the on-axis blur spot size of $0\fdg07$, which
was determined using another CCD measurement. Their larger spot size
is possibly due to saturation of the CCD, but it is no longer possible
to confirm this. Both spot sizes have reproduced shape parameter
distributions of observed background events reasonably, but they make
a significant difference in the narrower \textit{width} distributions
of simulated gamma-ray events which are displaced by about
$0\fdg02$. This is probably the main reason for the differences of the
optimized gamma-ray selection criteria based on the simulations
between this work and \citet{dazeley01a}.
\begin{deluxetable}{lcc}
\tablecaption{Comparison of simulation details between this work and
\citet{dazeley01a}. \label{tab_simulation_dp}}
\tablewidth{0pt}
\tablehead{
\colhead{Input Condition} & \colhead{This Work} & \colhead{\citet{dazeley01a}}
}
\startdata
EAS generation code
& GEANT 3.21  & MOCCA 92                \\
Composition
& p 100\%     & p 58\%, He 32\%, N 10\% \\
Power-law index of the integral energy spectrum
& $-1.7$      & $-1.65$                 \\
Zenith angle at the injection point
& $15\degr$   & $25\degr$               \\
Mirror reflectivity
& 75\%        & 66\%                    \\
On-axis blur spot size\tablenotemark{a}
& $0\fdg04$   & $0\fdg07$               \\
NSB rate\tablenotemark{b}
& 10\,MHz/PMT & 8.6\,MHz/PMT            \\
Signal loss
& 0\%         & 35\%                    \\
\enddata
\tablenotetext{a}{Standard deviation of the two-dimensional Gaussian.}
\tablenotetext{b}{Rate of photoelectron detections with the recoated mirror.}
\end{deluxetable}

\end{document}